\documentclass[
     amsmath,amssymb,
     aps,
    prd,
    floatfix,
    onecolumn,
    showkeys,
    12pt
    ]{revtex4-2}
    
    \usepackage{graphicx}
    \usepackage{bm}
    \usepackage{hyperref}
    \usepackage{amsmath} 
    \usepackage[version=4]{mhchem} 
    \usepackage[dvipsnames]{xcolor}

\usepackage{bm,nicefrac}

\usepackage{gensymb} 

\usepackage{subcaption}
\usepackage{cleveref}

\DeclareMathAlphabet{\mathbf}{OT1}{cmr}{bx}{n}

\usepackage{color}

\newcommand{\be}{\begin{equation}}
\newcommand{\ee}{\end{equation}}
\newcommand{\bea}{\begin{eqnarray}}
\newcommand{\beas}{\begin{eqnarray*}}
\newcommand{\eea}{\end{eqnarray}}
\newcommand{\eeas}{\end{eqnarray*}}
\newcommand{\ba}{\begin{array}}
\newcommand{\ea}{\end{array}}

\begin{document}

\title{Dirac CP phases in a 3+1 neutrino scenario with $\mu-\tau$ symmetry}
\author{Eduardo Becerra-Garc\'ia} \email{ebecerra@fis.cinvestav.mx} 
\author{Abdel P\'erez-Lorenzana}\email{aplorenz@fis.cinvestav.mx}
\affiliation{Departamento de F\'isica, Centro de Investigaci\'on y de Estudios Avanzados del I.P.N.\\ Apdo. Post. 14-740, 07000, Ciudad de M\'exico, M\'exico}

\begin{abstract}
A sterile neutrino in the $3+1$ scheme, where the sterile accounts for neutrino anomalies not explained solely by the weak active neutrinos, arises as a natural  source for the breaking of the $\mu-\tau$ symmetry suggested by oscillation neutrino data. We explore the predictions for the Dirac CP phases in this scenario, with and without sterile neutrino decay, and  show that current limits on $\delta_{CP}$ suggest a normal hierarchy and a lightest neutrino scale below 0.1~eV as the most plausible explanation for that, when Majorana phases are null. Other Dirac phases turn out to be non zero as well. 
\end{abstract}

\date{\today}
\maketitle


\section{Introduction}

Although the Standard Model (SM) has proved to be a very precise theory to explain most phenomena regarding fundamental particles and their interactions, it is by no means a complete theory, in the sense that there are still some issues for which it has not a solution, at least not in its original form, i.e., as a quantum field theory with the gauge group $SU(2)_L \times U(1)_Y$. One of these issues is related to the flavor sector of the SM, which introduces a numerous quantity of free parameters, an undesirable quality for a fundamental theory. If we could give a theoretical explanation for the value of those parameters, maybe we would be in the right direction in the search of a more fundamental theory. From the theoretical point of view that is one of the reason to study flavor symmetries, which relates the various generations of quarks and leptons, and which seem to be respected by all interactions described by the model, but for Yukawa interactions which are  responsible for providing masses to fundamental fermions.

Furthermore, in the SM, neutrinos are predicted to be massles particles, nonetheless, neutrino oscillations experiments, measuring neutrino fluxes that come from the sun, from cosmic ray  interactions at the upper atmosphere and from human made sources  including nuclear reactors and particle accelerators, have provided evidence beyond doubt that neutrinos are rather massive and  change their flavor as they travel from source to detector, a phenomena known as flavor oscillation. This last can be comprehended from  the fact that flavor neutrinos are not the actual mass eigenstates, but rather an admixture of them. 

In the three standard neutrino scheme, both flavor and mass neutrino basis are related by a unitary mixing matrix, that in the PMNS parametrization~\cite{Pontecorvo1957,Maki1962} is written in terms of three successive rotations, as $U_{mix} = R(\theta_{23})U(\theta_{13},\delta)R(\theta_{12})K$, where $R(\theta_{ij})$ is the real rotation in the $ij$ plane, with the angle $\theta_{ij}$. $U(\theta_{13},\delta)$ is a complex rotation on the $13$ plane, with an angle $\theta_{13}$ and   a Dirac phase $\delta$, and  $K$ is a diagonal matrix containing two Majorana phases. Oscillation probability along travel distance then becomes a function of the squared mass differences $\Delta m^2_{ij} = m_i^2-m_j^2$, for $i,j = 1,2,3$,  and mixing angles, through mixing matrix elements.
Global fit analysis of most oscillation data~\cite{Esteban2020} indicate that  
$\sin^2\theta_{12} = \sin^2\theta_{\odot}=0.304^{+0.012(3)}_{-0.012}$, $\sin^2\theta_{23}= \sin^2\theta_{ATM}=0.573(5)^{+0.016}_{-0.020(19)}$,  and  $\sin^2\theta_{13}=\sin^2\theta_{reac}= 0.02219(38)^{+0.00062(3)}_{-0.00063(2)}$, 
corresponding to solar, atmospheric and reactor oscillation mixings, respectively, for the normal (inverted) hierarchy of the mass spectrum. Here,  normal (inverted) refers to whether  (or not) $m_3^2$ is larger than $m_1^2$.  Due to MSW matter effects within the sun, the sign of the mass squared difference that governs solar neutrino oscillations is well defined, and  one gets  $\Delta m_{21}^2= \Delta m_{\odot}^2 = 7.42^{+0.21}_{-0.20}\times 10^{-5}~eV^2$. In contrast, as  the hierarchy is still unknown, 
$\Delta m^2_{ATM}$ corresponds to $\Delta m^2_{31}$  for normal hierarchy (NH),  or to $\Delta m^2_{23}$ for inverted hierarchy (IH). Its measured value as obtained from the same fits is $\Delta m_{ATM}^2 = 2.517^{+0.026}_{-0.028} (-2.498^{+0.028}_{-0.028})\times 10^{-3}~eV^2$.
Dirac CP phase comes to  $\delta_{CP}=197^{+27}_{-24}$ ($282^{+26}_{-30}$) and within the interval $[120,369]$ ($[193,352])$ at three sigma level.

The observation that the reactor mixing  is very small compared to other ones, and that $\sin^2\theta_{ATM}$ appears to be close to $1/2$, has suggested  an approximate $\mu-\tau$ exchange symmetry in the neutrino sector~\cite{Fukuyama1999,*Mohapatra1999,*Lam2001,*Ma2001} which implies that $|(U_{mix})_{\mu i}| =|(U_{mix})_{\tau i}|$. As a matter of fact, such a symmetry emerges explicitly when a null $\theta_{13}$ and a maximal $\theta_{23}$ are used to reconstruct the neutrino mass matrix from mass eigenvalues. Because of this, $\mu-\tau$ symmetry has been widely studied  as an appealing flavor symmetry for the sector
(for some earlier works see~\cite{Harrison2002,*Babu2002,*Ohlsson2002,*Kitabayashi2003,*Grimus2003,*Koide2004,*Mohapatra2004,*Ma2004,*Goshal2004,Mohapatra2006,*Joshipura2007,*Fuki2007,*Riazuddin2007,*Luhn2007,*Koide2008,*Honda208,*Ishimori2008,*Merle2014,*Rivera2016,Gomez2008,Rivera2015}. An extensive review can be found in \cite{Xing2016}).
$\mu-\tau$ symmetry, however, is at most an approximated one, because neither $\theta_{13}$ is consistent with zero, nor $\theta_{23}$ is maximal, not even at three sigma level. The source of the breaking of the symmetry is unknown, but it can be encoded within mass matrix elements as corrections to symmetry conditions that account for the final values of neutrino oscillation parameters (for flavor models  exploring the realization of the mixing matrix see for instance~\cite{Rahat:2018sgs,Perez:2019aqq}). 

Above mentioned results, though, do ignore data coming from LSND~\cite{lsnd} and MiniBooNe~\cite{miniboone2010,*miniboone2018} experiments, which have observed some events identified as the appearance above background of electron neutrinos (antineutrinos)  within an originally muon neutrino (antineutrino) flux. This observations are troublesome for the three standard neutrino scheme, since to explain such events within the picture of flavor oscillations, data implies a  larger mass scale, about $\Delta m^2 \sim 1~eV^2$,  with a rather small mixing angle, $\sin^2\theta_{LSND}\approx 0.0023$. To account for these, 
a fourth neutrino needs to be added, which does not participate from weak interactions. Such a sterile neutrino, $\nu_s$, has also been  motivated  by the discovery of the reactor antineutrino anomaly~\cite{Mueller2011,*Mention2011,*Huber2011}, associated to a deficit of the  reactor antineutrino detection rate in several experiments, and by the gallium neutrino anomaly~\cite{Abdurashitov2006,*Lavede2007,*Giunti2007b}. With  the presence of a sterile neutrino, the most favorable arrangement of the spectrum would be that of the $3+1$ hierarchy~\cite{Giunti2011,Giunti2019}, where all other neutrino oscillation experiments are explained by the lighter three states. Global fits with $3+1$ scheme, nonetheless, have shown that a serious tension exist among LSND/MiniBooNE and other short base line experiment appearance data when results from  disappearance experiments are included~\cite{Giunti2019,Dentler2018,hagstotz202}, that excludes the sterile solution up to 4.5 sigmas when the analysis is done without the low energy MiniBooNE data. This tension can be alleviated, although not solved, if the sterile is allowed to fast decay into electron neutrinos~\cite{Palomares2005} such that the appearance excess can be explained with smaller mixings, inducing a smaller effect on disappearance channels~\cite{Moulai2020,Diaz20201}. 
As shown in the  analysis of Ref.~\cite{Moulai2020}, that includes a one year data from IceCube, the global fit with the $3+1$ scenario with sterile neutrino decay provides a 2.8$\sigma$ improvement over the simple $3+1$ model predictions, with the best fit values $\Delta m^2\sim 1.35~eV^2$, $|U_{e4}| = 0.238$, $|U_{\mu 4}| = 0.105$, and a sterile lifetime $t = 4.5~eV^{-1}$.
New data regarding searches for a sterile neutrino, and the confirmation (or refutal) of mentioned anomalies, is expected along the forthcoming years from ongoing and planned short base line neutrino experiments. Meanwhile, the possible existence of an $eV$ scale sterile neutrino seems to stand up (for a recent review of current sterile search results see~\cite{Boser2020}).  

If it were to exist, sterile neutrinos can play a special roll in the understanding of the flavor neutrino problem. Since it does not participate of the weak interactions, sterile carries no standard flavor, and thus, it comes natural for it to violate flavor symmetries. This observation suggest in particular that the sterile could be the source for the breaking of the apparent $\mu-\tau$ symmetry we observe from oscillation data. This idea was already explored in a previous work~\cite{Rivera2015}, but no CP phases were considered in there. 
Our goal here is to revise the idea and explore possible implications of it for CP violation in the sector. For that, we shall assume that all the relevant violation of the $\mu-\tau$ symmetry comes from the mass terms that involve the sterile neutrino, whereas the active sector does obey the $\mu-\tau$ symmetry. High energy flavour model realizations of this idea were  explored in Refs.~\cite{Borah:2016fqj,Sarma:2018bgf} were some general hints on Dirac phase correlations were given for some scenarios (see also~\cite{Das:2018qyt,Das:2019ntw} for similar ideas). Besides the preference for the normal hierarchy, which is also hinted to from cosmology \cite{Vagnozzi2017}, we will also see a significant difference for the allowed mass values reported in this work when compared with the results obtained in \cite{Rivera2015}. Such distinction is due to the complex nature of the mass matrix elements in the sterile sector. Our results also improve those of~\cite{Borah:2016fqj,Sarma:2018bgf} on the search for solutions that are consistent with the expected PMNS CP phase.
The present  work is organized as follows.
In section 2 we briefly describe our active-sterile neutrino ansatz and the method of approximated diagonalization we use to calculate the mixings. Some important results and the system of equations that connect neutrino oscillation observables with mass matrix parameters is also presented there. In section 3 we explore the solutions to the system. In particular, we found the parameter space for the CP phases in terms of the lightest neutrino mass, in both normal and inverted hierarchy, for null Majorana phases, within the $3+1$ and $3+1+decay$ models. In section 4 we provide a brief discussion on the rephasing invariants. In section 5 we present a brief analysis for the case when the Majorana phases are not null in order to explore the sensitivity of our results to these last. We conclude in section 6 with some perspectives and outlook of this work. An appendix with a more detailed description of the diagonalization of the mass matrix has been added.


\section{Active-sterile neutrino mass matrix and oscillation parameters}

Under the hypothesis that active neutrino mass terms, $m_{\ell\ell^\prime} \bar{\nu}_\ell^c\nu_{\ell^\prime}$, for $\ell,\ell^\prime = e,\mu,\tau$, are invariant upon the discrete $\nu_\mu\leftrightarrow\nu_\tau$ exchange, the mass matrix elements  are forced to satisfy that $m_{e\mu} = m_{e\tau}$, and $m_{\mu\mu}= m_{\tau\tau}$, reducing the independent matrix elements to four. Furthermore, it can be shown that in such a case CP phases can all be factored out, and so, no CP violation is implied by the symmetry in the mixing matrix ($\delta_{CP} = 0$). 
The four remaining real parameters just account for the solar mixing and the three mass eingenvalues. This is easily seen  when we reconstruct the active neutrino mass matrix starting from the PMNS mixing matrix, setting $\theta_{13} = 0$ and $\theta_{23} = -\pi/4$, which gives the following relations among the parameters
\bea
m_{ee}&=& m_{1}\cos^{2}\theta_{12}+m_{2}\sin^{2}\theta_{12}~,\nonumber\\
m_{e\mu}&=&\frac{\sin2\theta_{12}}{\sqrt{8}}\left(m_{2}-m_{1}
\right)~,\nonumber\\
m_{\mu\tau}&=&\frac{1}{2}\left(m_{1}\sin^2\theta_{12}+m_{2}\cos^2\theta_{
12} -m_ { 3 } \right)~,\nonumber\\
m_{\mu\mu}&=&
\frac{1}{2}\left(m_{1}\sin^{2}\theta_{12}+m_{2}\cos^2 \theta_{12} 
+m_{3}\right)~.\label{mterms}
\eea
The last are easily inverted in favor of masses, as
\bea
m_{1}&=&m_{ee}-\sqrt{2}{m}_{e\mu}\tan\theta_{12},\nonumber \\
m_{2}&=&m_{ee}+\sqrt{2}{m}_{e\mu}\cot\theta_{12},\nonumber \\
m_{3}&=&{m}_{\mu\mu}-m_{\mu\tau}~,\label{masseigen}
\eea
where the solar to be $\theta_{12}$ mixing is given by
\be
\tan2\theta_{12}=
\sqrt{8}\left[\frac{{m}_{e\mu}}{{m}_{\mu\mu}+\left(m_{\mu\tau}-m_{ee}
\right)}\right]~.\label{12mix}
\ee

As already stated, the just described $\mu-\tau$ symmetry in the standard neutrino sector is at best an approximated symmetry, because reactor mixing is not null nor atmospheric one is maximal.
The problem then becomes to identify possible sources that produce the breaking. Of course, one possibility is that the same high energy mechanism that  generates neutrino masses be responsible for it. But it is also possible that the symmetry would be naturally violated in some other sector, and the effect communicated through interactions to weak flavor neutrinos. An example of the last is the mass difference among charged muon and tau leptons, which is far from zero. This violation to $\mu-\tau$ symmetry is, as a matter of fact, communicated to neutrino masses through weak charged interactions at one loop. Nevertheless, such induced mass correction turns out to be suppressed  by the $W$ mass and too small to account for the observed value of $\theta_{reac}$~\cite{Gomez2008}. 

Along this same line of thought, a natural candidate to look at is the sterile neutrino. If present, it would carry no standard flavor, but it has to couple to the weak flavours through mass terms.
It is then perfectly possible that the sterile sector should not comply with the symmetry. 
If sterile to active mass terms explicitly violate $\mu-\tau$, the  induced  effective mass corrections on the active sector would also break the symmetry. The effect should be expected to be  of the order of $m_\nu/m_s$, for $m_s$ the sterile mass, which should be just about what is needed to understand $\theta_{reac}$.  The idea  has been previously discussed in Ref.~\cite{Rivera2015}, where it has been shown that a sterile with a mass about $eV$ scale do allows for a successful reconstruction of mixings, within the known limitations of the sterile hypothesis,  although the analysis there was done without considering CP violation. Here we will revise this idea, including CP phases, to explore its implications, with particular interest on the possible predicted values for $\delta_{CP}$, and additional CP phases.

With a sterile neutrino, in the $3+1$ scheme where low scale neutrino oscillations are mainly explained by active mixings, whereas LSND/MiniBooNE is due to oscillations dominated by the sterile scale, the most general Majorana mass matrix that conserves $\mu-\tau$ symmetry in the weak flavor sector has the form 
\begin{equation}\label{Mmatrix}
    M_\nu =
    \begin{pmatrix}
        M_{S} & \bm{\alpha} m_s \hspace{1mm} \vspace{1mm} \\
        \bm{\alpha}^T m_s & m_s \hspace{1mm} \\
    \end{pmatrix},
\end{equation}
where $M_{S}$ is a $3\times 3$ $\mu-\tau$ symmetric mass matrix, which alone would produce the active neutrino masses given in Eq.~(\ref{masseigen}), howbeit, due to the sterile, we are now forced to maintain the former Majorana  phases of the two off-diagonal terms, $m_{e\mu}$ and $m_{\mu\tau}$ along the analysis.
In above, $m_s$ stands for the sterile neutrino mass and the vector $\bm{\alpha}^T  = (\alpha_e, \alpha_\mu, \alpha_\tau)$ represents the dimensionless and complex sterile to active neutrino mixing parameters, which by construction do not obey the symmetry and thus $\alpha_\mu\neq\alpha_\tau$. As already stated, in this scenario we are assuming that the couplings with the sterile neutrino would be the sole responsible for both, the deviation on $\mu-\tau$ predicted mixings, as well as for CP violation on the neutrino sector.

A small scale see-saw approximation, given by the decoupling of the sterile, shows that, at lower level, effective active neutrino mass terms would become  $m^\prime_{\rho \delta} \simeq m_{\rho \delta} - {\alpha}_{\rho}{\alpha}_{\delta}m_{s}$.  Thus, the breaking of the symmetry, as induced by the sterile to active mass mixings, can be encoded in the effective (and complex) parameters defined as $\delta = m^\prime_{e\tau} -  m^\prime_{e\mu}\simeq \alpha_e\,\Delta\alpha\, m_s$ and 
$\epsilon = m^\prime_{\tau \tau} - m^\prime_{\mu \mu} \simeq 
2\bar\alpha_\mu\,\Delta\alpha\, m_s$, where  
$\bar\alpha_\mu= (\alpha_\mu+\alpha_\tau)/2$,  and $\Delta\alpha= \alpha_\tau-\alpha_\mu$. Note that $\Delta\alpha$ is the only parameter that actually  measures the amount of breaking of the symmetry, and the one that would be responsible for the nonzero value of reactor mixing. $\alpha_e$, on the other hand, would have the role of fixing the initial value of $\theta_{12}$ to produce the solar mixing. CP conserving analysis has indicated that 
a consistent solution can be obtained from $\alpha's\approx {\cal O}(10^{-1})$~\cite{Rivera2015}. The addition of CP phases should slightly modify those results, but a positive solution should still be expected.
The further exploration of how this affects masses and mixings could be pursued in this small seesaw approximation.  Of course,  the limitations of such an approach resides in the decoupling of the sterile that deprive us from explicitly accounting for the sterile neutrino evidences themselves. To provide a more complete analysis, all masses and mixings derived from the complete mass matrix in Eq.~(\ref{Mmatrix}) should be taken into account. We shall follow this path hereafter, with the main aim of studying  the  phase space that is consistent with the expected oscillation neutrino parameters. The interest on this arises when we realize that $M_\nu$ is parametrized by at most thirteen parameters. Five real mass terms, four in $M_{S}$ plus $m_s$, two free phases therein, and three  $\alpha$ mixing parameters and their three phases. On the other hand, these parameters should account for sixteen physical observables in the sector, given by six mixing angles, three in the active sector and three active to sterile mixings, six CP phases (three Majorana and three Dirac like ones) and four mass eigenvalues. As it is clear, since there are less parameters than observables, there would have to be some definite predictions from the model. As most mixings and mass square mass differences are known, it is natural to  think that the possible predictions could better be seen in a so far less constrained sector, the CP phase space, that we now proceed to explore. To this aim, we shall next  diagonalize the neutrino mass matrix  and force its parameters to reproduce the observed oscillation parameters, and then,  move into exploring the remaining phase space, in particular the one associated to Dirac like CP phases.

By replacing the sterile parameters by $\alpha_\ell\rightarrow \alpha_\ell\, e^{i\phi_\ell}$, to explicitly express their CP phases,  the mass matrix $M_\nu$ is  written as
\begin{equation}
    M_{\nu}=
    \begin{pmatrix}
        m_{ee} & m_{e\mu} & m_{e\mu} & \alpha_e e^{i\phi_e} m_s \hspace{1mm} \\
        m_{e\mu} & m_{\mu\mu} & m_{\mu\tau} & \alpha_\mu e^{i\phi_\mu} m_s \hspace{1mm} \\
        m_{e\mu} & m_{\mu\tau} & m_{\mu\mu} & \alpha_\tau e^{i\phi_\tau} m_s \hspace{1mm} \\
        \alpha_e e^{i\phi_e} m_s & \alpha_\mu e^{i\phi_\mu} m_s & \alpha_\tau e^{i\phi_\tau} m_s & m_s \hspace{1mm}\\ 
    \end{pmatrix}.
\end{equation}
This matrix can be diagonalized by the unitary transformation $U^T M_\nu U = M_D$, with $U$ the four by four mixing matrix defined through the relation 
$\nu_{\alpha L} = \sum_{i} U_{\alpha i} \nu_{i L}$
where hereafter $i \in \{1,2,3,4\}$ and $\alpha \in \{e,\mu,\tau,s\}$. $U$ can be parametrized in terms of six mixing angles and six phases. These last would be in general non trivial expressions involving  the  parameters of $M$ and the $\phi_\ell$ phases. Those are the result of the diagonalization process and the parametrization used for $U$. Nevertheless, we should emphasize that, for the purpose of our study, it is not necessary to display those expressions explicitly, but rather knowing the relation of the physical phases with all other observables, as we show below.
In this work we use the `symmetrycal parametrization' of the mixing matrix proposed by J. Schechter and J.W.F. Valle in Ref.~\cite{Schechter1980},
\begin{equation}
\label{U_mix}
U = \omega_{34}(\theta_{34},\gamma) \omega_{24}(\theta_{24},\beta) \omega_{14}(\theta_{14},\alpha) \omega_{23}(\theta_{23},\delta_3) \omega_{13}(\theta_{13},\delta_2) \omega_{12}(\theta_{12},\delta_1),
\end{equation}
with the $\omega_{ij}$ matrices given by
\begin{equation}
\label{U4}
    [\omega_{ij}(\theta_{ij}, \phi)]_{pq} =
    \begin{cases}
        \cos{\theta_{ij}} &\quad p=q=i,j,\\
        1 &\quad p=q\neq i,j,\\
        \sin{\theta_{ij}} \hspace{1mm} e^{-i\phi} &\quad p=i,q=j,\\
        -\sin{\theta_{ij}} \hspace{1mm} e^{i\phi} &\quad p=j,q=i,\\
        0	&\quad \text{otherwise.}
    \end{cases}
\end{equation}
In this parametrization, three phases can be factored out and recombined into the Majorana phases, but there are three phase combinations that cannot be extracted from $U$, and they would become 
the physical Dirac phases appearing in the oscillation probabilities. Following Ref.~\cite{Rodejohann2011}, these are
\begin{align}
    I_1 &= \delta_1+\delta_3-\delta_2, \label{CPphase1}\\
    I_2 &= \delta_1+\beta-\alpha,\label{CPphase2} \\
    I_3 &= \delta_2+\gamma-\alpha, \label{CPphase3}
\end{align}
 where $I_1$ corresponding to the $\delta_{CP}$ appearing in the PMNS mixing matrix. $I_{2,3}$, on the other hand, are additional  phases for which we still do not have experimental inputs.

We block diagonalize the mass matrix (see Appendix) following a procedure analogous to the one presented in Ref.~\cite{King2002}, assuming the angles $\theta_{14}, \theta_{24}, \theta_{34}$ and $\theta_{13}$ to be small enough such that a perturbative approach is justified. As stated in the Appendix, along the process we use a systematic step by step redefinition of neutrino phases that warrants the real value of the  obtained mixings and amounts to define the phases of $U$ as given in Eq.~(\ref{U_mix}). Those results, formally, should allow us to express the more useful mixing matrix parameters in terms of the mass matrix elements. We will use the former in the rest of our discussion.

In order to make the connection of the proposed ansatz with the experimental data, we invert the diagonalization and rewrite it as $M = U^* M_D U^{\dagger}$. From this relation we obtain the following approximate relations, valid for either mass ordering, between the mixing angles of the sterile sector and the $\alpha$ parameters of the mass matrix,
\begin{align}
\theta_{14} &\approx \alpha_e,
\label{theta14_alpha} \\
\theta_{24} &\approx \alpha_\mu, \\
\theta_{34} &\approx \alpha_\tau. 
\label{theta34_alpha}
\end{align}
This expressions can also be obtained from the approximated diagonalization of the mass matrix, as discussed in the Appendix, in the limit of sterile mass dominance (upon phase redefinitions).

The general expression for the transition probability  is given by
\begin{align}
    P_{\nu_\alpha \rightarrow \nu_\beta}(L,E) = \delta_{\alpha\beta} &- 4\sum_{k>j} Re[U^*_{\alpha k}U_{\beta k}U_{\alpha j}U^*_{\beta j}]\sin^2{\Bigg(\frac{\Delta m^2_{kj}L}{4E}\Bigg)} \nonumber \\ &+ 2\sum_{k>j} Im[U^*_{\alpha k}U_{\beta k}U_{\alpha j}U^*_{\beta j}]\sin{\Bigg(\frac{\Delta m^2_{kj}L}{2E}\Bigg)},
    \label{pstd}
\end{align}
from where, the survival probability becomes
\begin{equation}
    P_{\nu_\alpha \rightarrow \nu_\alpha}(L,E) = 1 - 4\sum_{k>j} |U_{\alpha k}|^2 |U_{\beta k}|^2\sin^2{\Bigg(\frac{\Delta m^2_{kj}L}{4E}\Bigg)}.
\end{equation}
Due to the hierarchy of the neutrino masses in the $3+1$ scenario, the above expressions can be simplified as follows,
\begin{align}
    P_{\nu_\alpha \rightarrow \nu_\beta} \approx &  A^{\alpha\beta}_{LSND} \sin^2{\Delta_{LSND}} + A^{\alpha\beta}_{ATM} \sin^2{\Delta_{ATM}} + A^{\alpha\beta}_{\odot} \sin^2{\Delta_{\odot}} \nonumber \\ &+ B^{\alpha\beta}_{ATM} \sin{2\Delta_{ATM}} + B^{\alpha\beta}_{\odot} \sin{2\Delta_{\odot}}
\end{align}
for the transition probabilities (for $\alpha\neq\beta$), and
\begin{align}
    P_{\nu_\alpha \rightarrow \nu_\alpha} \approx 1 - A^{\alpha}_{LSND}  \sin^2{\Delta_{LSND}} - A^{\alpha}_{ATM} \sin^2{\Delta_{ATM}} - A^{\alpha}_{\odot}  \sin^2{\Delta_{\odot}}
\end{align}
for the survival probabilities, where the amplitudes are given by 
\begin{eqnarray*}
    A^{\alpha\beta}_{LSND} &=& 4|U_{\alpha 4}|^2|U_{\beta 4}|^2,  \\
    A^{\alpha\beta}_{ATM} &=& 4|U_{\alpha 3}|^2|U_{\beta 3}|^2 + 4 Re (U_{\alpha 3}^* U_{\beta 3} U_{\alpha 4} U_{\beta 4}^*),  \\
    A^{\alpha\beta}_{\odot} &=& -4 Re (U_{\alpha 2}^* U_{\beta 2} U_{\alpha 1} U_{\beta 1}^*), \\
    A^{\alpha}_{LSND} &=& 4|U_{\alpha 4}|^2(1-|U_{\alpha 4}|^2),\\
 A^{\alpha}_{ATM} &=& 4|U_{\alpha 3}|^2(|U_{\alpha 2}|^2+|U_{\alpha 1}|^2),\\
    A^{\alpha}_{\odot} &=& 4|U_{\alpha 1}|^2|U_{\alpha 2}|^2,\\
    B^{\alpha\beta}_{ATM} &=& -2 Im(U_{\alpha 3}^* U_{\beta 3} U_{\alpha 4} U_{\beta 4}^*), \\
    B^{\alpha\beta}_{\odot} &=& 2 Im(U_{\alpha 2}^* U_{\beta 2} U_{\alpha 1} U_{\beta 1}^*), 
\end{eqnarray*}
and the oscillation arguments given by the neutrino scales $\Delta_{LSND} = \Delta m^2_{LSND} L/4E$, $\Delta_{ATM} = \Delta m^2_{ATM} L/4E$, and $\Delta_{\odot} = \Delta m^2_{\odot} L/4E$. Note that in general, all three CP violating Dirac phases, $I_{1,2,3}$, would be involved in the amplitude coefficients, in particular those entering the appearance probabilities.
The effective mixing angles associated to different types of neutrino oscillations (solar, reactor, atmospheric and LSND/MiniBooNE) can be read out from above amplitudes, considering the two neutrino interpretation of each experiment, and so, they can be written as
\begin{align}
   \sin^2{2\theta_\odot} & \approx 4|U_{e1}|^2|U_{e2}|^2,\label{obs_mix_sol} \\
   \sin^2{2\theta_{reac}} & \approx 4|U_{e3}|^2(|U_{e1}|^2+|U_{e2}|^2), \\
   \sin^2{2\theta_{ATM}} & \approx 4|U_{\mu 3}|^2|U_{\tau 3}|^2 
   + 4 Re (U_{\mu 3}^* U_{\tau 3} U_{\mu 4} U_{\tau 4}^*), \\
   \sin^2{2\theta_{LSND}} & \approx 4|U_{e4}|^2|U_{\mu 4}|^2.\label{obs_mix_4}
\end{align}
It is worth noticing that the effective form of solar and reactor mixings is the same as in the standard three neutrino flavor, in terms of mixing matrix elements, whereas only the atmospheric mixing gets and explicit correction coming from the sterile sector.

With the  use of the parametrization given in (\ref{U_mix}) for the mixing matrix, one straightforwardly gets the following explicit expressions for the required  mixing matrix elements
\begin{eqnarray*}
U_{e1} &=& \cos\theta_{14}\cos\theta_{13}\cos\theta_{12},        \\
U_{e2} &=& \cos\theta_{14}\cos\theta_{13}\sin\theta_{12}e^{-i\delta_1}, \\
U_{e3} &=& \cos\theta_{14}\sin\theta_{13}e^{-i\delta_2},     \\
U_{e4} &=& \sin\theta_{14}e^{-i\alpha},        \\
U_{\mu 3} &=& \cos\theta_{24}\cos\theta_{13}\sin\theta_{23}e^{-i\delta_3} - \sin\theta_{14}\sin\theta_{24}\sin\theta_{13}e^{i(\alpha -\beta -\delta_2)},  \\
U_{\mu 4} &=& \cos\theta_{14}\sin\theta_{24}e^{-i\beta},  \\
U_{\tau 4} &=& \cos\theta_{24}\cos\theta_{14}\sin\theta_{34}e^{-i\gamma}  
\end{eqnarray*}
and
\begin{align*}
U_{\tau 3} =& \cos\theta_{34}\cos\theta_{13}\cos\theta_{23} -
 \sin\theta_{34}\cos\theta_{24}\sin\theta_{14}\sin\theta_{13} e^{i(\alpha -\gamma -\delta_2)}\\
& - \sin\theta_{34}\sin\theta_{24}\sin\theta_{23}\cos\theta_{13} e^{i(\beta -\gamma -\delta_3)} ~.
\end{align*}
Thus, with the results in (\ref{theta14_alpha}-\ref{theta34_alpha}), the right-hand side of Eqs.~(\ref{obs_mix_sol}-\ref{obs_mix_4}) can be written, up to second order in the small mixing angle $\theta_{13}$, and $\alpha_{e,\mu,\tau}$, as
\begin{align}
    \sin^2{2\theta_\odot} \approx &\, 4(1-2\theta_{13}^2-2\alpha_e^2)\sin^2{\theta_{12}}             \cos^2{\theta_{12}} \label{sin_thetasol},\\
    \sin^2{2\theta_{reac}} \approx &\, 4\theta_{13}^2, \label{theta_reac} \\
    \sin^2{2\theta_{ATM}} \approx &\, 4[ (1-2\theta_{13}^2 -\alpha_\mu^2-\alpha_\tau^2) \cos{\theta_{23}}
    \nonumber\\
    &-
    2\alpha_\mu \alpha_\tau \cos(\beta - \delta_3 -\gamma)\sin{\theta_{23}} ] \sin^2{\theta_{23}} \cos{\theta_{23}} \nonumber\\
    &+4\alpha_\mu \alpha_\tau \cos(\beta-\delta_3-\gamma)\sin\theta_{23}\cos\theta_{23},\label{sintheta_atm}\\
    \sin^2{2\theta_{LSND}} \approx &\, 4\alpha_e^2 \alpha_\mu^2.
    \label{sin_thetaLSND} 
\end{align}
As expected, the above expressions reduce to the standard ones in the limit when there is no sterile, i.e., with $\alpha_{e,\mu,\tau}$=0.

Since we have imposed $\mu-\tau$ symmetry to the active sector, then we have another pair of equations, $m_{e\mu}=m_{e\tau}$ and $m_{\mu\mu}=m_{\tau\tau}$, which in terms of the parameters of the mixing matrix and the mass eigenvalues are given by
\begin{align}
    \begin{split}
    &(-m_1 e^{-i\delta_1} + m_2 e^{i\delta_1})(\cos{\theta_{23}}+e^{-i \delta_3}\sin{\theta_{23}})\sin{\theta_{12}}\cos{\theta_{12}} \\
    &+ m_3 e^{i\delta_2}\theta_{13}(e^{i\delta_3}\sin{\theta_{23}}-\cos{\theta_{23}}) + m_4 e^{i\alpha} \alpha_e (\alpha_\mu e^{i\beta}-\alpha_\tau e^{i\gamma}) \approx 0,
    \end{split}
    \label{eqsym1}
\end{align}
\begin{align}
    \begin{split}
    &(m_1 e^{-2i\delta_1}\sin^2{\theta_{12}} + m_2 \cos^2{\theta_{12}})(\cos^2{\theta_{12}} - e^{-2i\delta_3}\sin^2{\theta_{12}})\\
    &+ m_3(e^{2i\delta_3}\sin^2{\theta_{12}} -\cos^2{\theta_{12}}) + m_4 (\alpha_\mu^2 e^{2i\beta} - \alpha_\tau^2 e^{2i\gamma}) \approx 0.
    \label{eqsym2}
    \end{split}
\end{align}
Notice that above relations, which are simply derived from mass matrix diagonalization, are valid regardless whether sterile neutrino decay is considered as part of the explanation of LSND/Miniboone data, but for $\sin\theta_{LSND}$,
 because the decay mainly affects the residual contributions of the heavy state to lighter neutrino state oscillations.
To be clear, with sterile neutrino decay, the transition probability calculation has to separate the simple mixing contributions to the amplitude coming from standard (and stable) neutrino flavors from those of the heavier (unstable) state. Thus, the transition probability has  to be written as
\begin{equation}
    P_{\nu_\alpha\rightarrow\nu_\beta} = P^{STD}_{\nu_\alpha\rightarrow\nu_\beta} +P^{s}_{\nu_\alpha\rightarrow\nu_\beta}
    +P^{decay}_{\nu_\alpha\rightarrow\nu_\beta}~,
\end{equation}
where $P^{STD}_{\nu_\alpha\rightarrow\nu_\beta}$ is given by the standard oscillation formula (\ref{pstd}) including only the lighter states. 
On the other hand, sterile contributions come now with an exponential suppression over distance in the amplitude probability, due to the decay, that gives
\begin{equation}
P^{s}_{\nu_\alpha\rightarrow\nu_\beta} =
 |U_{\alpha 4}|^2 |U_{\beta 4}|^2 \left(e^{-L/2t}- 1\right) + 2Re \sum_{j}^3  U^*_{\alpha 4} U_{\beta 4} U_{\alpha j} U^*_{\beta j} \left(e^{-2i\Delta_{LSND}}e^{-L/2t} - 1\right)~,
\end{equation}
where $t$ stands for the life time of the sterile. 
$P^{decay}_{\nu_\alpha\rightarrow\nu_\beta}$, on the other hand, provides the contribution that comes from the probability that the heavy component of a neutrino of flavor $\alpha$ decays at a shorter distance than $L$ and its product interacts as a $\beta$ neutrino.
The explicit form of this contribution has no relevance for the identification of the effective oscillation mixings, though.  A detailed discussion of this contribution can be found in Ref.~\cite{Palomares2005}.

Considering above arguments, even for the  sterile decay scenario, Eqs.~(\ref{sin_thetasol}-\ref{sintheta_atm}) remain the same due to unitarity of the mixing matrix. Although the formula (\ref{obs_mix_4}) can still be used for our proposes as a reference, given that the global fits with sterile decay give values for $|U_{e4}|$ and $|U_{\mu 4}|$,  we will use these last instead for this case.

The left hand side in Eqs.~(\ref{sin_thetasol}-\ref{sin_thetaLSND}) is known from experiment, and  the mass eigenvalues that appear in Eqs.~(\ref{eqsym1}, \ref{eqsym2}) can be expressed in terms of the squared-mass differences and a neutrino absolute scale parameter, $m_0$, which by convention is given as the lightest neutrino mass, which means to write 
\begin{equation}
    m_1 = m_0, \quad |m_2|=\sqrt{m_0^2+\Delta m_{\odot}^2}, \quad |m_3| = \sqrt{m_0^2+|\Delta m_{ATM}^2|}, \quad \text{for NH, and}
\end{equation}
\begin{equation}
|m_1| = \sqrt{m_0^2+|\Delta m_{ATM}^2|}, \quad |m_2|=\sqrt{m_0^2++|\Delta m_{ATM}^2+\Delta m_{\odot}^2}, \quad m_3 = m_0, \quad \text{for IH,} \\
\end{equation}
and $\Delta m^2_{LSND/MiniBooNE} \approx m_4^2$.
Therefore, by using this data as inputs for our analysis,  the set of eight real  Eqs.~(\ref{sin_thetasol}-\ref{eqsym2}) gets defined by twelve unknowns, given by the  three active mixing angles and six phases used in the parametrization (\ref{U4}) aside to the three $\alpha_\ell$ parameters. Without any further knowledge of other observable parameters, we are forced to make some reasonable assumptions on the unknowns in order to explore the parameter space in a more comprehensive way.
First we focus on exploring the Dirac like phase space, and thus restrict our study to
the particular case of null Majorana phases, i.e., we set $\alpha = \delta_1 = \delta_2 = 0$. The case for other non-zero values of these phases shall be considered later on. We consider these parameters as the Majorana phases since they are the ones appearing in the effective mass $|m_{ee}|^2$ of neutrinoless double beta decay, whose maximum value corresponds to the null values of such phases. Then, the relevant set of phases  in (\ref{CPphase1}-\ref{CPphase3}) is restricted to $\delta_3$, $\beta$ and $\gamma$. 
In the parametrization we are using and with the previous consideration, the phase $\delta_3$ corresponds to the usual Dirac phase $\delta_{CP}$ that appears in the PMNS  parametrization. Furthermore, following Eq.~(\ref{sin_thetaLSND}), we notice that $\alpha_e$ and $\alpha_\mu$ are tightly correlated,  so we can use some given values that comply with the perturbative condition $\alpha_{e,\mu}\ll 1$ to be able to depict the phase parameter space on two dimensional graphs, as we explain next. 


\section{Dirac CP phases}

As stated above, we seek for solutions of the system of equations (\ref{sin_thetasol}-\ref{eqsym2}) using the known neutrino oscillation parameter data, as reported in \cite{Esteban2020}, as constraints on the parameter space of our model.
For the analysis  
we use  for the sterile sector two previously known results, corresponding to the $3+1$ and $3+1+decay$ scenarios. For the first one we use $\sin^2{2\theta_{LSND}} \approx 0.0023$, $m_4 = 1~eV$ according to \cite{Giunti2011}, whereas for the second one we take
the best fit values $|U_{e4}|= 0.238$ and $|U_{\mu 4}|= 0.105$, meaning  $\sin^2{2\theta_{LSND}} \approx 0.0025$,  and $m_4 = 1.35~eV$ according to \cite{Moulai2020}.
We also use the best fit values for $\Delta m_{\odot}^2$ ($7.42\times 10^{-5}eV^2$ for both hierarchies), $\Delta m_{ATM}^2$ ($2.517\times 10^{-3}eV^2$ for NH, $-2.498\times 10^{-3}eV^2$ for IH), $\theta_{reac}$ ($8.57\degree$ for NH, $8.6\degree$ for IH), and  $\theta_{\odot}$ ($33.44\degree$ for NH, $33.45\degree$ for IH).
Also, we explore our remnant parameter space using 
a linear distribution in the range $[0, 0.2]~eV$ for the lightest neutrino scale $m_0$, for any given hierarchy. 

\begin{figure}[h!]
  \centering
    \includegraphics[width=0.45\linewidth]{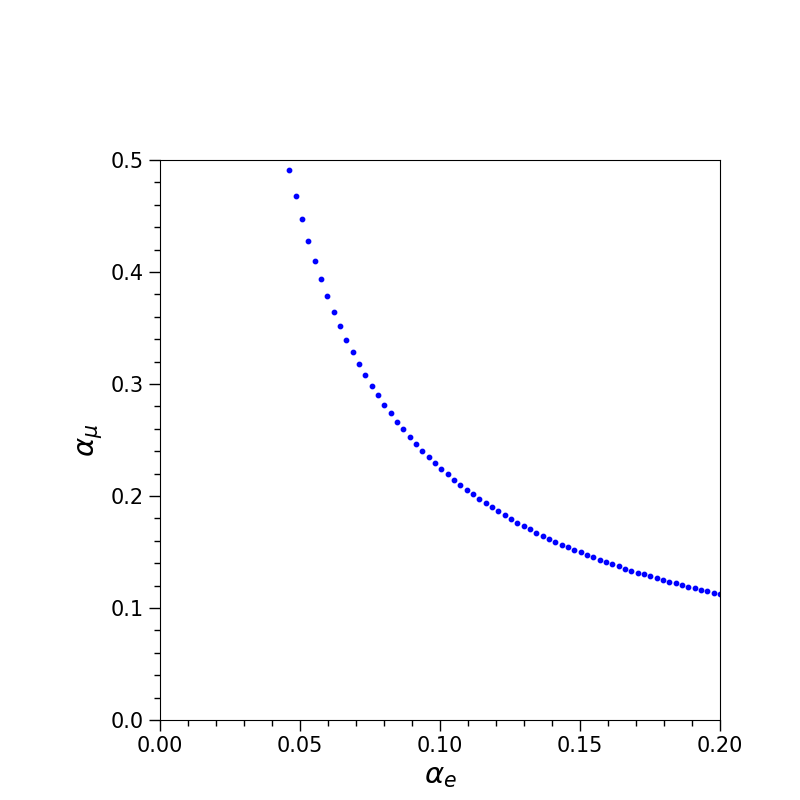}
    \caption{Allowed values for $\alpha_e$ and $\alpha_\mu$ for the best fit value of $\sin^2\theta_{LSND}$, as discussed in the text.}
    \label{alphas_sol}
\end{figure}

Next, we observe that, according to Eq.~(\ref{theta_reac}), $\theta_{reac}\approx\theta_{13}$, which fixes the value of this parameter. 
The values for $\alpha_{e,\mu}$ that solve Eq.~(\ref{sin_thetaLSND}), in the 3+1 scenario, and which are of interest in this work are shown in figure \ref{alphas_sol}. Choosing a pair of values from this plot allows us to solve Eq.~(\ref{sin_thetasol}) for $\theta_{12}$.
In the sterile decay scenario, $\alpha_e$ and $\alpha_\mu$ are fixed by the best fit values from Ref.~\cite{Moulai2020} that give $\alpha_e\approx|U_{e4}|  =  0.238$ and $\alpha_\mu\approx|U_{\mu 4}|= 0.105$, same that we take for such a case along the analysis. With such values for $\alpha_{e,\mu}$, Eq.~(\ref{sin_thetasol}) has no real solution for $\theta_{12}$ with the best fit value for $\theta_\odot$, but we can certainly find a set of solutions for sligthly smaller values of $\theta_{\odot}$ lying within the $1\sigma$ interval.
We are left to determine which values of the following five unknown parameters, $\theta_{23}$, $\alpha_\tau$, $\beta$, $\gamma$, and $\delta_{CP}$ (which is given by $\delta_3$ as already mentioned)
provide the right value for $\theta_{ATM}$  satisfying the four real conditions derived from  Eqs.~(\ref{eqsym1}) and (\ref{eqsym2}). For this purpose, since 
Eq.~(\ref{sintheta_atm})  already suggests that $\theta_{23}$ should be close to the actual value of $\theta_{ATM}$, we proceed to numerically explore the five parameter space by allowing $\theta_{23}$ angle to vary within the ($3\sigma$) interval [$40.1\degree,  51.7\degree$] for NH and [$40.3\degree , 51.8\degree$] for IH, and solve all five  constraining conditions to determine the remaining parameter values that are consistent with current neutrino mixing observables (up to 3$\sigma$ level in $\theta_{ATM}$), with the main goal of determining the potential predictions for $\delta_{CP}$.

\begin{figure}[h!]
  \centering
  \begin{subfigure}[b]{0.45\linewidth}
    \includegraphics[width=\linewidth]{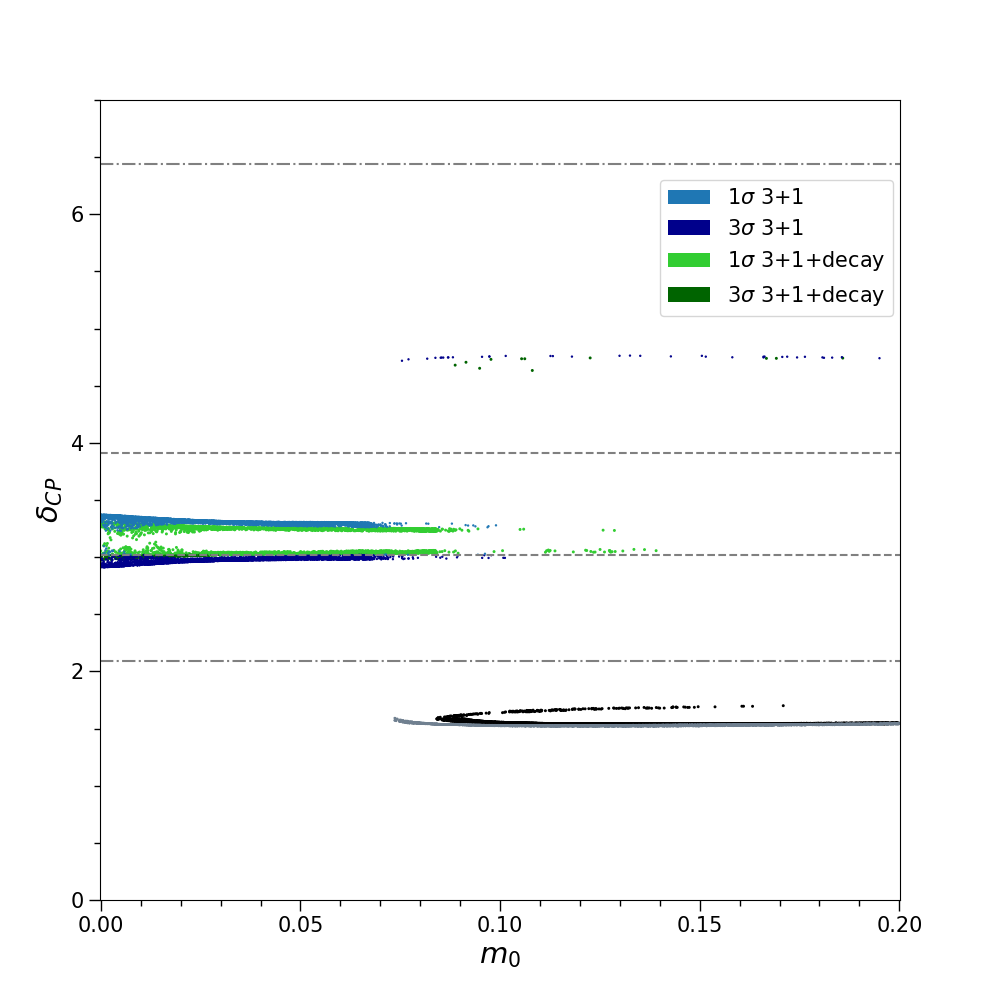}
    \caption{Normal hierarchy.}
    \label{delta_m0_NH_2}
  \end{subfigure}
  \begin{subfigure}[b]{0.45\linewidth}
    \includegraphics[width=\linewidth]{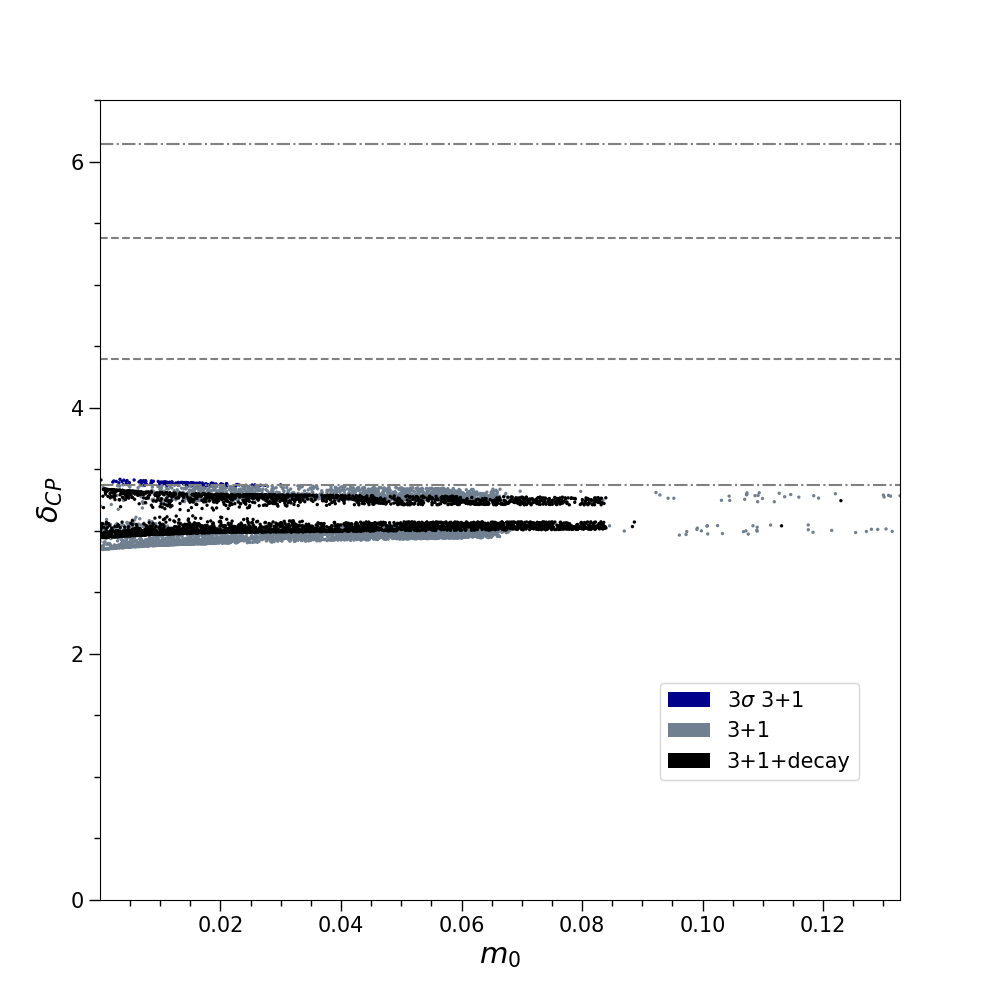}
    \caption{Inverted hierarchy.}
    \label{delta_m0_IH_2}
  \end{subfigure}
  \caption{Predictions for the Dirac phase $\delta_{CP}$ as a function of the lightest neutrino mass $m_0$ for normal (a) and inverted (b) hierarchy. For the $3+1$ scenario light (dark) blue dots lie within the current $1\sigma$ ($3\sigma$) region represented by the horizontal lines, analogously for the $3+1+decay$ scenario. Gray and black dots lie outside the $3\sigma$ range.}
  \label{delta3_m0}
\end{figure}

In figure \ref{delta3_m0} we show the results for $\delta_{CP}$ as a function of $m_0$ for appropriate values of $\alpha_e$ and $\alpha_\mu$ as required in Ref.~\cite{Rivera2015}. The horizontal dashed and dashdotted lines indicate the $1\sigma$ ($3.0194\rightarrow 3.9095$ for NH, $4.3982\rightarrow 5.3756$ for IH) and $3\sigma$ ($2.0944\rightarrow 6.4402$ for NH, $3.3685\rightarrow 6.1436$ for IH) current intervals for the Dirac CP phase $\delta_{CP}$, respectively. The plots presented in this section correspond to $\alpha_e \approx 0.1910$, $\alpha_\mu \approx 0.1176$, 
which imply that  $\theta_{12} \approx 39.03\degree$
for the $3+1$
scenario, which is one solution to Eqs.~(\ref{sin_thetasol}) and (\ref{sin_thetaLSND}). We have explored solutions for different values of the parameters $\alpha_e, \alpha_\mu$, and found only small changes in the results for the whole parameter space. Therefore we show here only the given set of  solutions to visually represent the main conclusions of the whole analysis and comment on the effect of changing the values of $\alpha_{e,\mu}$ on a parameter case basis. In fact, with a larger value for $\alpha_e$ (hence smaller for $\alpha_\mu$), it is possible to reproduce very similar results to those obtained in the $3+1+decay$ scenario, as expected, since these values would be near the best fits already mentioned. 
We found a notorious difference between the normal and inverted hierarchies for the phase $\delta_{CP}$. In the first case we see there are solutions lying in the $1\sigma$ as well as in the $3\sigma$ range, these are presented in figure~\ref{delta_m0_NH_2} for the $3+1$ ($3+1+decay$) scenario with light blue (light green) and dark blue (dark green) dots, respectively, while gray (black) dots correspond to solutions of the system of equations that are outside the $3\sigma$ interval. The same color map is applied to all the plots presented in this section and the next one.

\begin{figure}[h!]
  \centering
  \begin{subfigure}[b]{0.45\linewidth}
    \includegraphics[width=\linewidth]{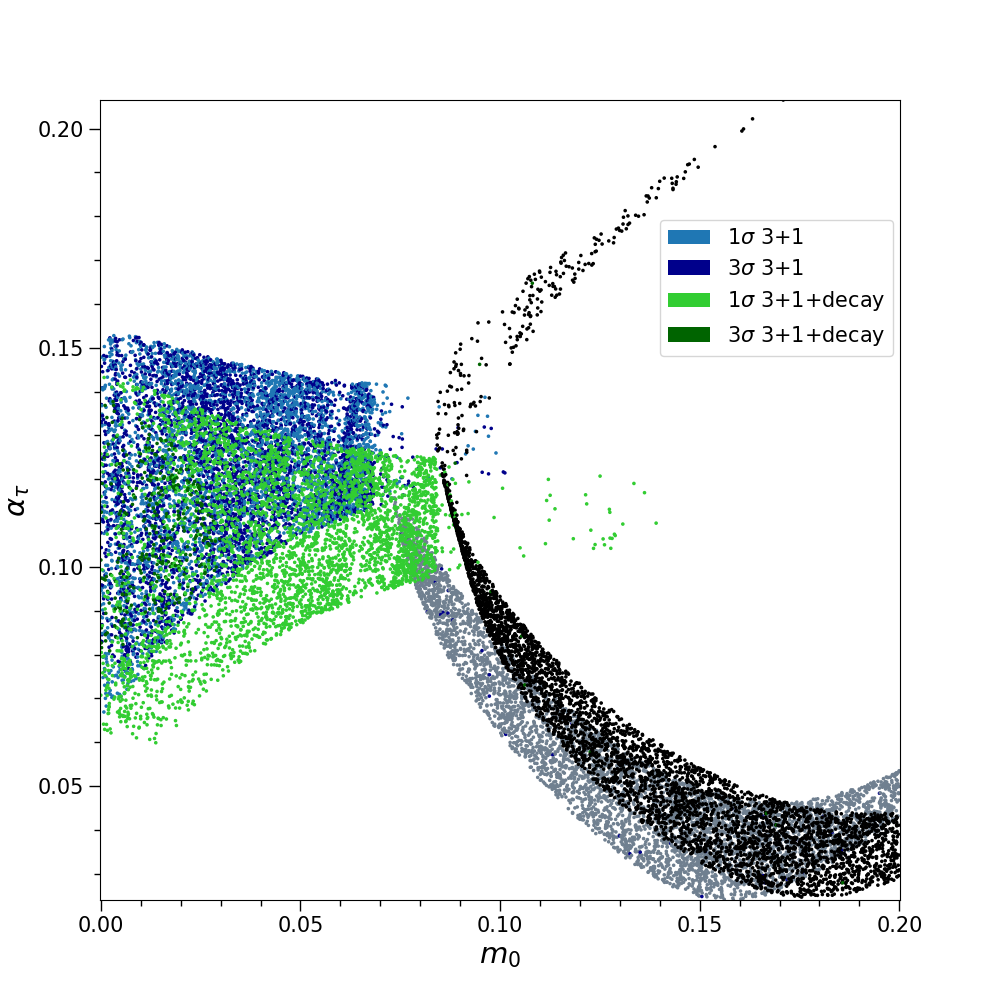}
    \caption{Normal hierarchy.}
    \label{alpha_m0_NH_2}
  \end{subfigure}
  \begin{subfigure}[b]{0.45\linewidth}
    \includegraphics[width=\linewidth]{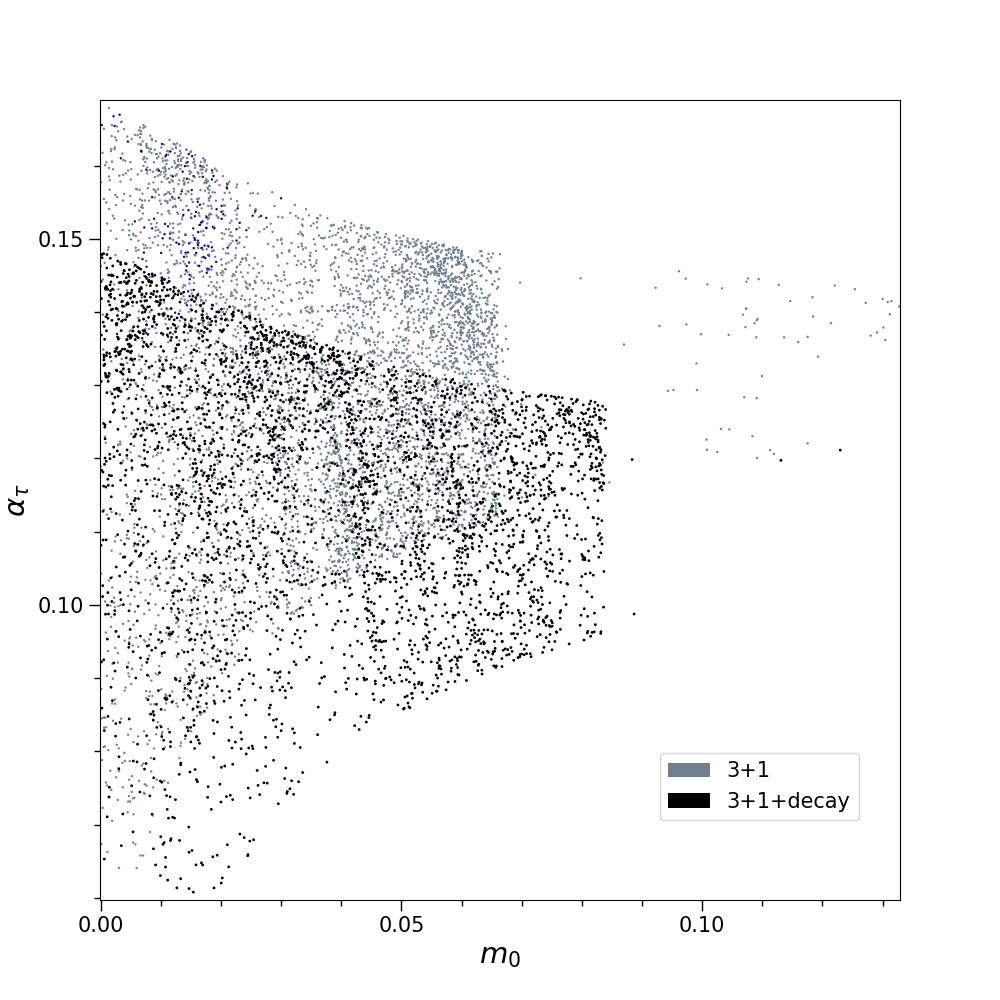}
    \caption{Inverted hierarchy.}
    \label{alpha_m0_IH_2}
  \end{subfigure}
  \caption{Parameter space for $\alpha_\tau$, consistent with neutrino oscillation data,  as a function of the lightest neutrino mass $m_0$, corresponding to  $\alpha_e \approx 0.1910$, $\alpha_\mu \approx 0.1176$  and $\theta_{12} \approx 39.03\degree$ for the $3+1$ scenario, and for the $3+1+decay$ scenario as discussed in the text.}
  \label{alpha_tau}
\end{figure}

Note that the vast majority of predicted points within the $1\sigma$ and $3\sigma$ intervals correspond to solutions for which the mass $m_0$ is below $0.075eV$, favoring a hierarchical neutrino spectrum, and we see from figure \ref{alpha_m0_NH_2} that the corresponding $\alpha_\tau$ values lie in the approximated interval $(0.07,0.15)$ for the NH, but there is not any clear differentiation among $1\sigma$ and $3\sigma$ sets. For larger (smaller) values of $\alpha_e$, the gap between the two regions in figure \ref{delta_m0_NH_2} decreases (increases) which implies there are more (fewer) points in the $1\sigma$ region, but in any case the mass $m_0$ remains below $0.1eV$. The region for $\alpha_\tau$ changes slightly remaining within the allowed perturbative values. The results contrast with those found in Ref.~\cite{Rivera2015}, where the lightest neutrino mass was required to be greater than $0.1eV$ (with no CP violation and without decay). Hence the allowed parameter space for the neutrino mass is drastically changed by the presence of the phases. For the inverted hierarchy we only obtained a few solutions within the $3\sigma$ interval as can be seen in figure \ref{delta_m0_IH_2}, indicating that the analysis favours the normal hierarchy according to the present experimental data. The few dark blue points in this case are also below $0.1eV$, as in the NH case. For other values for $\alpha_{e,\mu}$ we obtained similar results, with even fewer points lying in the $3\sigma$ region.

\begin{figure}[h!]
  \centering
  \begin{subfigure}[b]{0.45\linewidth}
    \includegraphics[width=\linewidth]{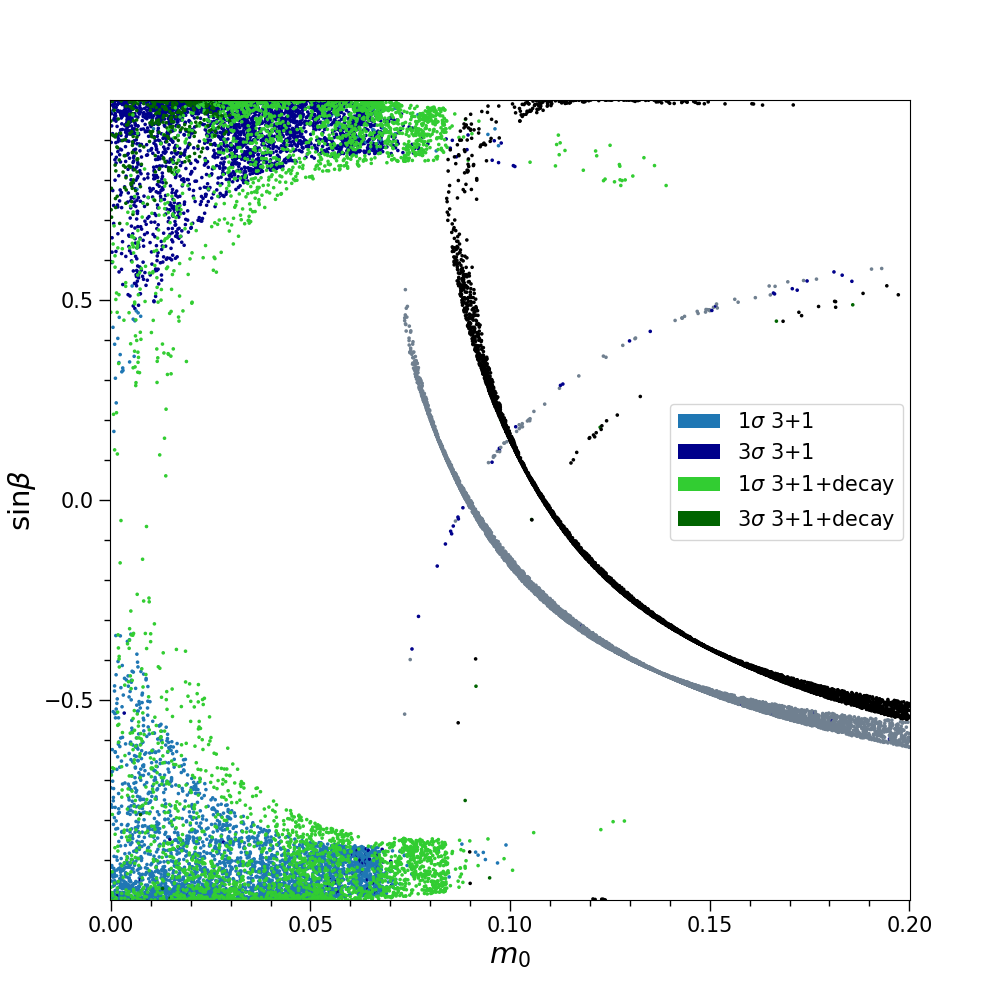}
    \caption{Normal hierarchy.}
  \end{subfigure}
  \begin{subfigure}[b]{0.45\linewidth}
    \includegraphics[width=\linewidth]{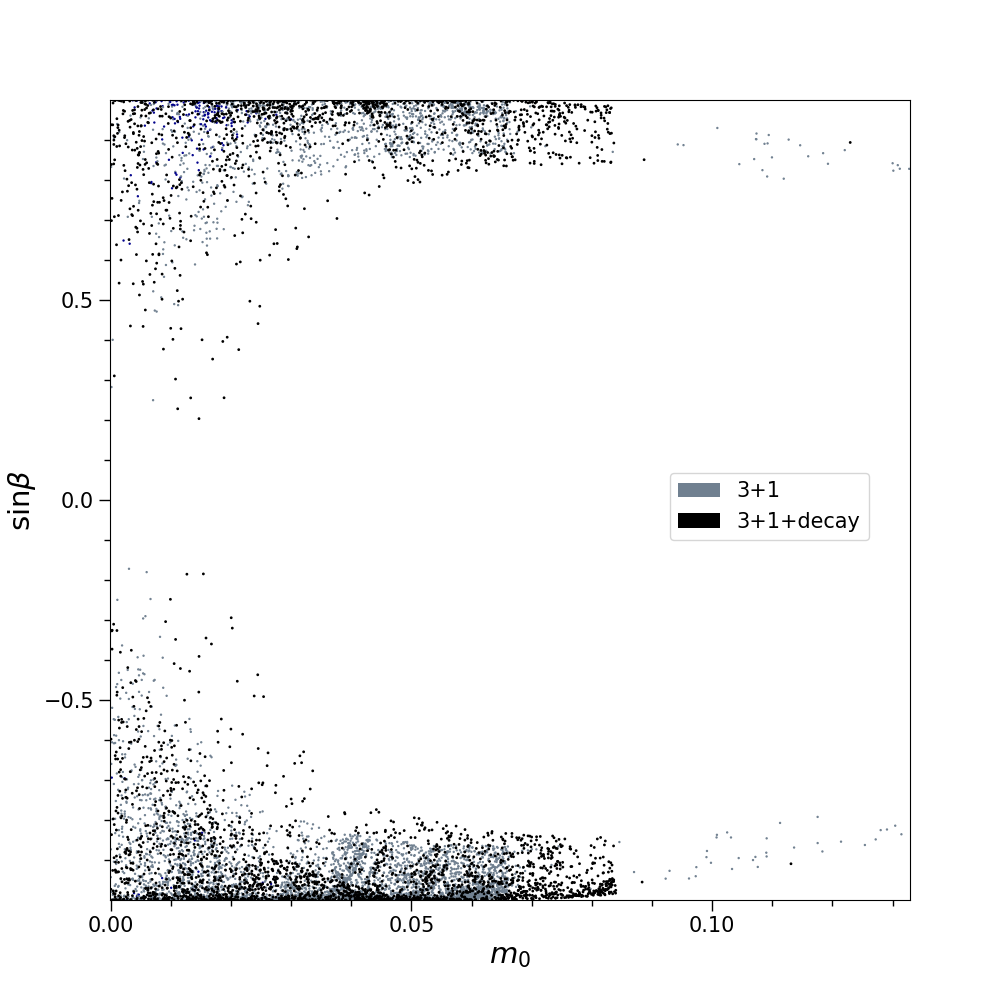}
    \caption{Inverted hierarchy.}
  \end{subfigure}
  \caption{$\sin{\beta}$ as a function of the lightest neutrino mass $m_0$. The majority of the positive values for NH correspond to the $3\sigma$ interval for $\delta_{CP}$ while the negative ones are within $1\sigma$.}
  \label{sinbeta_m0}
\end{figure}

\begin{figure}[h!]
  \centering
  \begin{subfigure}[b]{0.45\linewidth}
    \includegraphics[width=\linewidth]{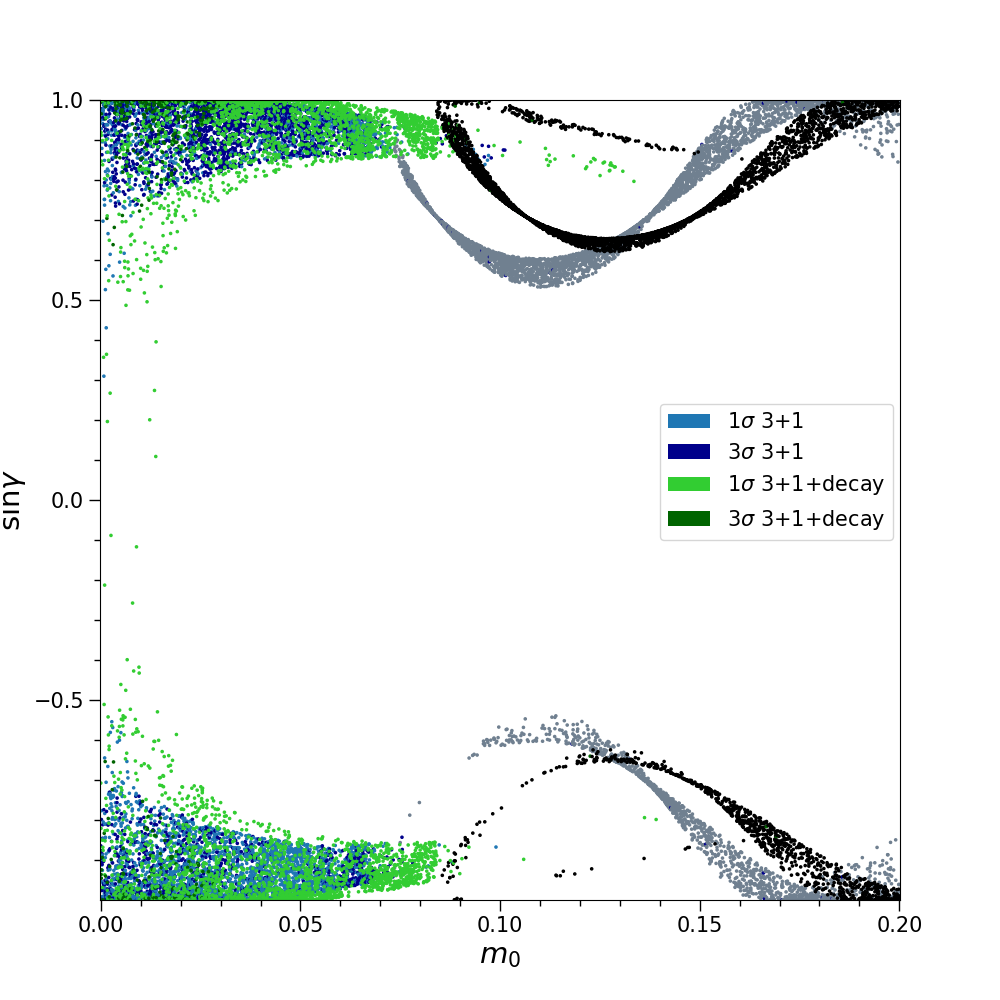}
    \caption{Normal hierarchy.}
  \end{subfigure}
  \begin{subfigure}[b]{0.45\linewidth}
    \includegraphics[width=\linewidth]{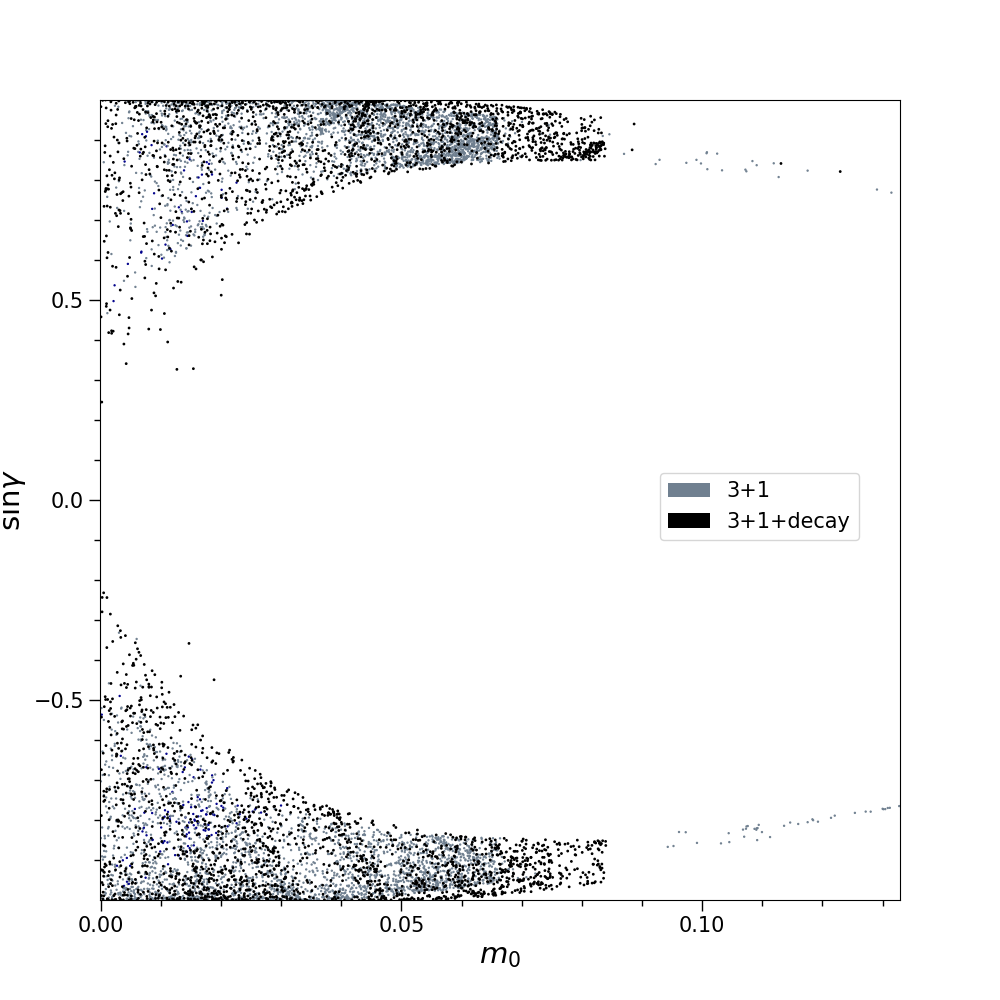}
    \caption{Inverted hierarchy.}
  \end{subfigure}
  \caption{$\sin{\gamma}$ as a function of the lightest neutrino mass $m_0$. The majority of the positive values for NH correspond to the $3\sigma$ interval for $\delta_{CP}$ while the negative ones are within $1\sigma$.}
  \label{singamma_m0}
\end{figure}

In figures \ref{sinbeta_m0} and \ref{singamma_m0} we present the results for the other two CP phases, $\beta$ and $\gamma$. Since $\alpha=\delta_1=\delta_2=0$, the Dirac phases  $I_2$ and $I_3$ in Eqs.~(\ref{CPphase2},\ref{CPphase3}) reduce to $\beta$ and $\gamma$, respectively. Therefore, these are the corresponding physical CP phases appearing in neutrino oscillations. We plot the sine of these parameters as a function of the lightest neutrino mass $m_0$. For the NH case we see a pronounced difference for positive and negative values of $\sin{\beta}$ in the sense that the negative values correspond to the solutions lying within the $1\sigma$ interval, while the positive values correspond to the $3\sigma$ interval. For the IH case, the majority of the solutions lying within the $3\sigma$ interval correspond to positive values of $\sin{\beta}$. This character of the plots is also true for the $\sin{\gamma}-m_0$ plots, as we can see in figure \ref{singamma_m0}. It is also remarkable that null values of these phases do not appear to be  consistent with the current bounds on $\delta_{CP}$. Larger values of $\alpha_e$, generate more positive solutions for $\sin{\beta}$ and $\sin{\gamma}$ within the $1\sigma$ region in NH, while smaller values distinguish even more the two regions. For the IH case, we found a similar behaviour of the plots for different values of $\alpha_{e,\mu}$ but always outside the $3\sigma$ region. Furthermore, note that a future improvement on the measurement of $\delta_{CP}$ would better constrain the parameter space for these extra phases, since each dot in these plots correspond to one single value of $\delta_{CP}$.

\section{Rephasing invariants}
One very convenient approach to the study of CP violation in the flavor sector of the standard model is by means of the rephasing invariants, since these are independent of the parametrization of the mixing matrix. In general, they are given by \cite{Giunti2007}
\begin{equation}
    \mathcal{J}_{\alpha \beta }^{kj} = Im[U_{\alpha k}U_{\beta j}U_{\alpha j}^*U_{\beta k}^*].
\end{equation}
Due to the unitarity of $U$ they are not all independent, for our case, with four neutrino mixing,  there are nine of such invariants, which we can identify as $\mathcal{J}_{\tau s}^{13}$, $\mathcal{J}_{\tau s}^{14}$, $\mathcal{J}_{\tau s}^{34}$, $\mathcal{J}_{se}^{13}$, $\mathcal{J}_{se}^{24}$, $\mathcal{J}_{se}^{34}$, $\mathcal{J}_{e\mu}^{23}$, $\mathcal{J}_{e\mu}^{24}$, $\mathcal{J}_{e\mu}^{34}$ according to Ref.~\cite{Guo2002}. Here we are interested in the invariant defined by the elements of the active sector, $\mathcal{J}_{e\mu}^{23}$, because it is the one that quantifies the CP violation in weak neutrino oscillations,  besides, it reduces to the Jarlskog invariant when we switch off the sterile sector, i.e., with $\theta_{i4} = 0$. This invariant can be written, with null Majorana phases, as follows,
\[
\mathcal{J}_{e \mu}^{23} = \left[ 
  s_{13} s_{14} s_{24} c_{12} c_{23}\sin{\beta} - 
  s_{23} c_{12} c_{13} c_{23} c_{24}\sin{\delta_{3}} -
s_{12} s_{14} s_{23} s_{24} \sin{(\beta - \delta_{3})}
\right]~ s_{12} s_{13} c_{13} c_{14}^2 c_{24}~,
\]
where, to simplify matters,  $c_{ij}$ ($s_{ij}$) stands for $\cos\theta_{ij}$ ($\sin\theta_{ij}$).

By writing this same expression  as a function of the  $\alpha_e$, $\alpha_\mu$, $\alpha_\tau$ parameters up to second order we have
\begin{equation}
\mathcal{J}_{e \mu}^{23} =
    \theta_{13} \left(- \alpha_{e} \alpha_{\mu} s_{12} \sin{\left(\beta - \delta_{3} \right)} + \left(0.5 \alpha_{e}^{2} + \alpha_{\mu}^{2} - 1\right)  c_{12} c_{23} \sin{\delta_{3}} \right) s_{12} s_{23}~.
\end{equation}
\begin{figure}[h!]
  \centering
  \begin{subfigure}[b]{0.45\linewidth}
    \includegraphics[width=\linewidth]{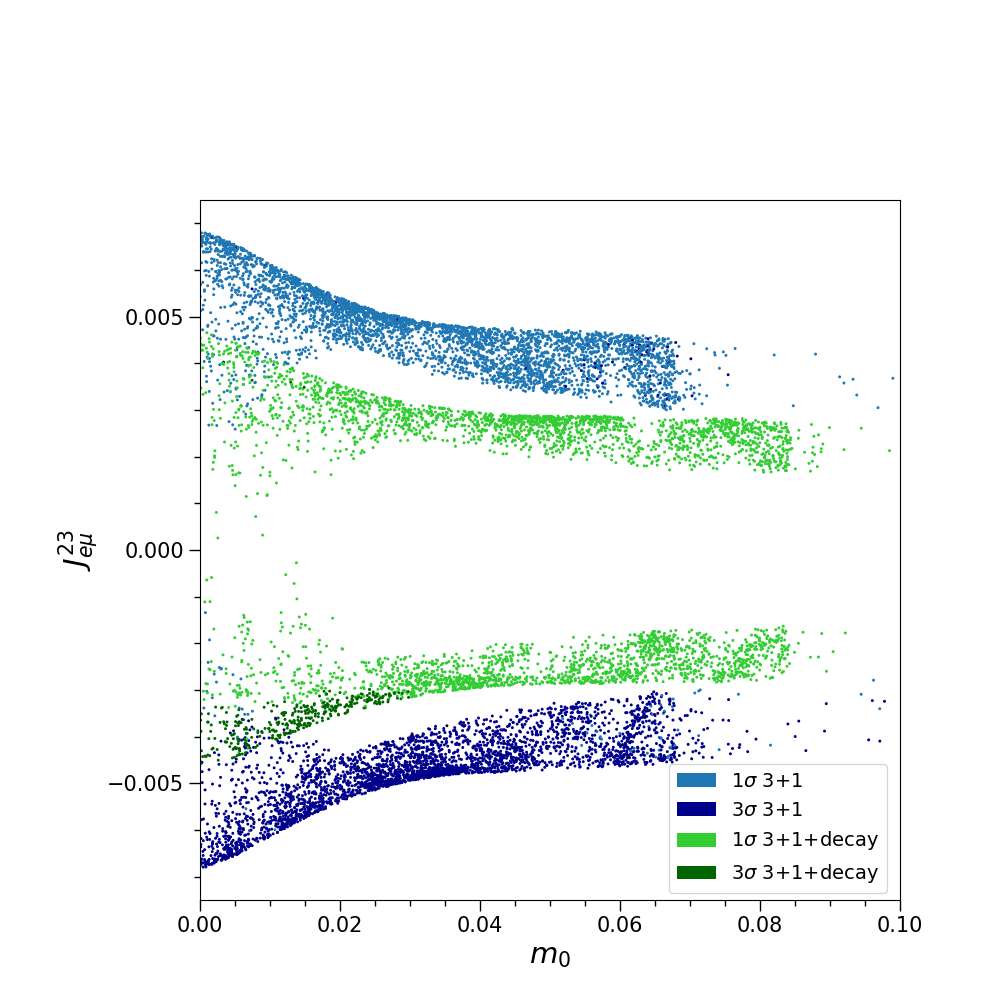}
    \caption{Normal hierarchy.}
  \end{subfigure}
  \begin{subfigure}[b]{0.45\linewidth}
    \includegraphics[width=\linewidth]{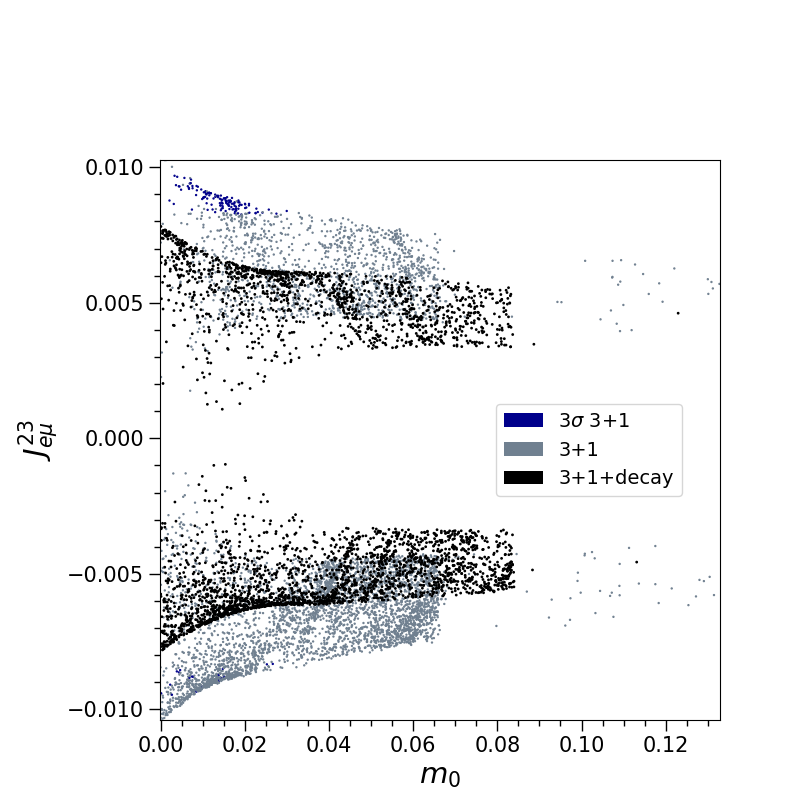}
    \caption{Inverted hierarchy.}
  \end{subfigure}
  \caption{Jarslkog rephasing invariant $J_{e\mu}^{23}$ as a function of the smallest neutrino mass for both hierarchies.}
  \label{invariant1}
\end{figure}

In figure \ref{invariant1} we plot the latter quantity as a function of the lightest neutrino mass $m_0$ for both normal and inverted hierarchies where the color map indicates the same attributes as before. As we would expect, there is a significant difference for the NH an IH scenarios. We see the same tendency for the solutions within $1\sigma$ and $3\sigma$ to be clearly distinguished for the NH case, in the $3+1$ scheme, the former are the mirror of the latter with respect to $J_{e\mu}^{23}=0$, while in the case with sterile decay, the majority of the solutions lie within the $1\sigma$ interval, hence such scenario is preferred for CP violation. For the IH case the majority of the points lie outside the $3\sigma$ interval of $\delta_{CP}$, as expected, since the invariant is a function of the phases.

Regarding the other invariants, up to second order in the $\alpha$ parameters and the small mixing angle $\theta_{13}$, we obtained the following expressions,
\begin{equation}
\mathcal{J}_{e\mu}^{24} \approx 
      \alpha_{e} \alpha_{\mu} \left[ \theta_{13} s_{12} s_{23} \sin{\left(\beta - \delta_{3} \right)} + \left(\frac{1}{2}~ \theta_{13}^{2} - 1\right)  c_{12} c_{23} \sin{\beta}\right] s_{12}~, 
\end{equation}
\begin{equation}
\mathcal{J}_{e\mu}^{34} \approx  
    - \alpha_{e} \alpha_{\mu} \theta_{13} s_{23} \sin{\left(\beta - \delta_{3} \right)}~,
\end{equation}
\begin{eqnarray}
\mathcal{J}_{\tau s}^{13}/ \cos{\theta_{23}} &\approx&
    \alpha_{e} \alpha_{\mu} \Bigg(\theta_{13}^{2} \left(\left(\frac{3}{2} s^{2}_{23} - 1\right)  s_{12} c_{12} \sin{\beta} - 2 s_{12} s^{2}_{23} c_{12} \cos{\left(\beta - \delta_{3} \right)}\sin{\delta_{3}}\right) \nonumber\\
    & +& \theta_{13} \left(2 s^{2}_{12} - 1\right) s_{23}c_{23} \sin{\left(\beta - \delta_{3} \right)} +  s_{12} s^{2}_{23} c_{12} \sin{\beta}\Bigg) \nonumber\\
    &+& \alpha_{e} \alpha_{\tau} \Bigg(\theta_{13} \left(1- 2 s^{2}_{23} ) s^{2}_{12} -  c^2_{23}\right)\sin{\gamma} \nonumber\\
    &+& \left(\theta_{13}^{2} \left( - \frac{1}{2} \sin{\delta_{3}} \cos{\gamma} + \frac{3}{2} \sin{\gamma} \cos{\delta_{3}}\right) + \sin{\left(\delta_{3} + \gamma \right)}\right) s_{12}c_{12} s_{23}  c_{23} \Bigg) \nonumber \\
    &+& \alpha_{\mu} \alpha_{\tau} \Bigg(\frac{1}{2} \theta_{13}^{2} s^{2}_{12} s_{23} \sin{\left( \delta_{3} + \gamma - \beta \right)} - \theta_{13} \left( s^{2}_{23} + c^{2}_{23}\right) s_{12}  c_{12} c_{23} \sin{\left(\beta - \gamma \right)} \nonumber\\
    &-& s^{2}_{12} s_{23} \sin{\left( \delta_{3} + \gamma - \beta \right)}\Bigg)
    + \theta_{13} \left(\alpha_{e}^{2} - \alpha_{\mu}^{2}\right)  s_{12} s_{23} c_{12} \sin{\delta_{3}}~,
\end{eqnarray}
\begin{eqnarray}
\mathcal{J}_{\tau s}^{14} &\approx&
    \alpha_{e}\alpha_{\tau} c_{12}\left(\theta_{13} c_{12}c_{23} \sin{\gamma}  - \sin{\theta_{12}} \sin{\theta_{23}} \sin{\left(\delta_{3} + \gamma \right)}\right) \nonumber\\
    &+& \alpha_{\mu}\alpha_{\tau} c_{23} \Big(\theta_{13} \left(\cos{(2\theta_{23})}\sin{\left(\beta - \gamma \right)} + 2  s^{2}_{23}s_{12} c_{12}\sin{\delta_{3}} \cos{\left( \delta_{3} + \gamma- \beta  \right)}\right)  \nonumber\\
    & +&\left( s^{2}_{12}-\theta_{13}^{2} c^{2}_{12} \right) s_{23} \sin{\left(- \beta + \delta_{3} + \gamma \right)}\Big)~, 
\end{eqnarray}
\begin{equation}
\mathcal{J}_{\tau s}^{34} \approx
    \alpha_{\tau} \left( \alpha_{\mu}\left(\frac{1}{2} \theta_{13}^{2} - 1\right) s_{23} \sin{\left( \delta_{3} + \gamma- \beta  \right)}-\alpha_{e} \theta_{13} \sin{\gamma}\right) c_{23}~,
\end{equation}
\begin{eqnarray}
\mathcal{J}_{se}^{13} &\approx&
    \theta_{13} \Big( \alpha_{e} \alpha_{\tau} \left(\theta_{13} s_{12} s_{23} \sin{\left(\delta_{3} + \gamma \right)} -  c_{12} c_{23} \sin{\gamma}\right) \nonumber\\ 
    &-&\alpha_{e} \alpha_{\mu}\left( \theta_{13}  s_{12} c_{23} \sin{\beta} + c_{12} s_{23} \sin{\left(\beta - \delta_{3} \right)} \right)     \nonumber\\
    &+& \alpha_{\mu} \alpha_{\tau} \left(2 s^{2}_{23}  \sin{\delta_{3}}\cos{\left( \delta_{3} + \gamma - \beta \right)} - \sin{\left(\beta - \gamma \right)}\right) s_{12}  \nonumber\\
    &+& \left( \alpha_{\tau}^{2}- \alpha_{\mu}^{2} \right) s_{12} s_{23} c_{23}\sin{\delta_{3}} \Big) c_{12}~,
\end{eqnarray}
\begin{eqnarray}
\mathcal{J}_{se}^{24} &\approx&
    \alpha_{e} s_{12} \Bigg(\alpha_{\mu} \left(\theta_{13} s_{12} s_{23} \sin{\left(\beta - \delta_{3} \right)} + \left(\frac{1}{2} \theta_{13}^{2} - 1\right)  c_{12} c_{23}\sin{\beta}\right) \nonumber\\
    &+& \alpha_{\tau} \left(\theta_{13} s_{12} c_{23}\sin{\gamma} + \left(1 - \frac{1}{2} \theta_{13}^{2}\right) s_{23} c_{12} \sin{\left(\delta_{3} + \gamma \right)} \right)\Bigg)~,
\end{eqnarray}
and
\begin{equation}
\mathcal{J}_{se}^{34} \approx
    - \alpha_{e} \theta_{13} \big(\alpha_{\mu} s_{23} \sin{\left(\beta - \delta_{3} \right)} + \alpha_{\tau} c_{23} \sin{\gamma}\big)~.
\end{equation}

\section{Switching on Majorana phases}

\begin{figure}[h!]
  \centering
  \begin{subfigure}[b]{0.45\linewidth}
    \includegraphics[width=\linewidth]{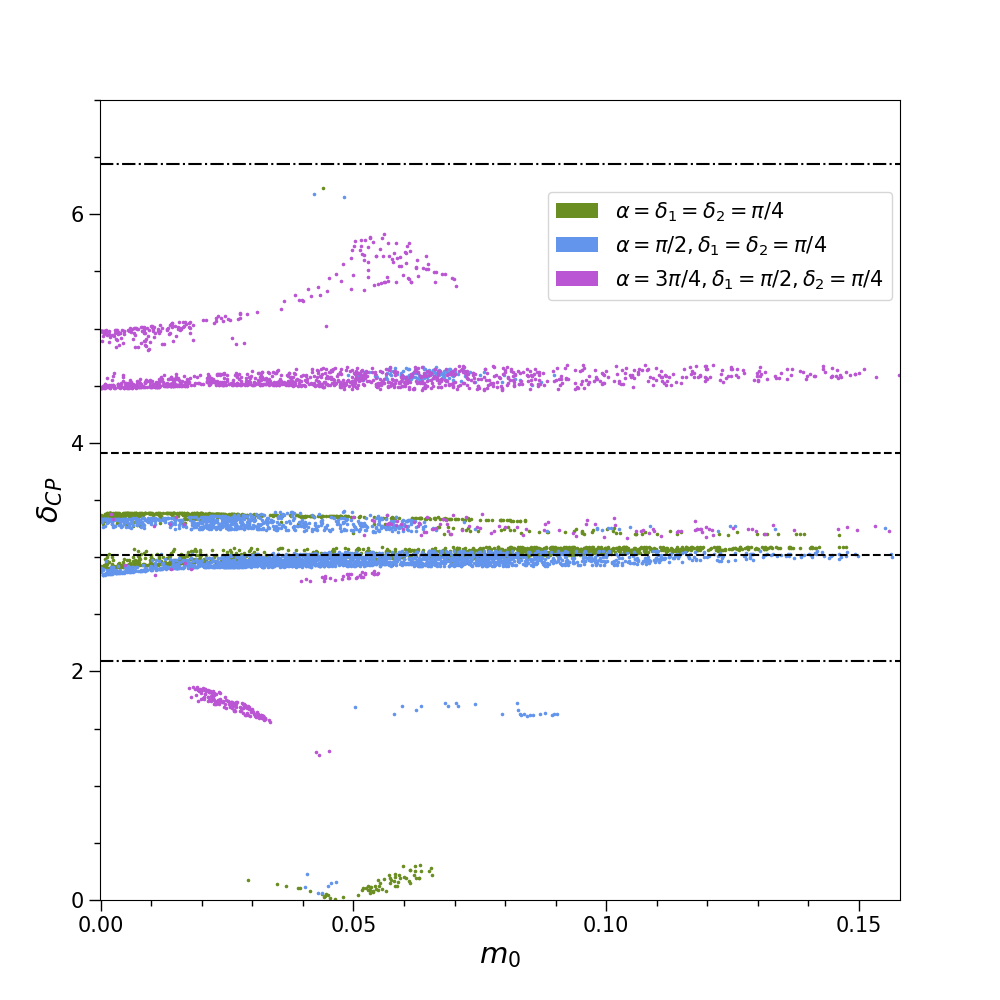}
    \caption{Normal hierarchy.}
  \end{subfigure}
  \begin{subfigure}[b]{0.45\linewidth}
    \includegraphics[width=\linewidth]{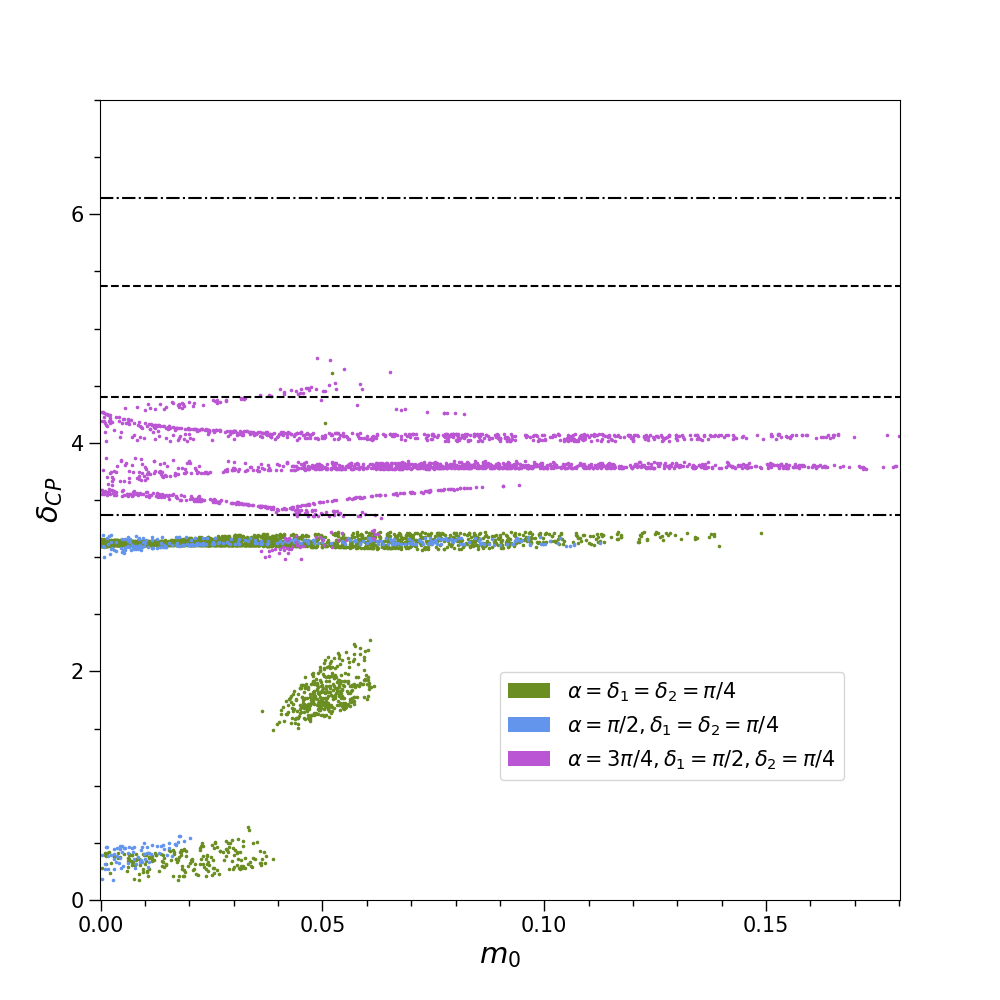}
    \caption{Inverted hierarchy.}
  \end{subfigure}
  \caption{Predictions for the Dirac phase $\delta_{CP}$, for the normal (a) and inverted (b) hierarchies, in three scenarios of non-zero Majorana phases. As before, the horizontal dashed and dashdotted lines indicate the $1\sigma$ and $3\sigma$ current intervals for the Dirac CP phase $\delta_{CP}$, respectively.}
  \label{delta3_mp}
\end{figure}

In this section we briefly discuss the case when the Majorana phases are not null, within the $3+1+decay$ model. First, because it is more consistent with data than the simple $3+1$ scenario, and second, since there is actually little visual difference among parameter spaces, as the results of section III had shown. Since we do not have any experimental data regarding these phases, we fix them to certain arbitrary values in order to explore how they affect the solutions previously reported. We study three cases, in the first one all of them are equal, (i) $\alpha = \delta_1 = \delta_2 = \pi/4$, in the second case two phases are equal, (ii) $\alpha = \pi/2$, $\delta_1=\delta_2 = \pi/4$, and in the third all three phases are fixed to different values, (iii) $\alpha = 3\pi/4$, $\delta_1= \pi/2$, $\delta_2 = \pi/4$. In the following plots, green, blue and purple dots represent the solutions for each of these cases, respectively. Since now we have non-zero Majorana phases, we have to take into account the relations (\ref{CPphase1}-\ref{CPphase3}), with $I_1$ corresponding to the usual $\delta_{CP}$ phase.
In figure \ref{delta3_mp} we show the parameter space for $\delta_{CP}$, and we see there are still solutions within the $1\sigma$ and $3\sigma$ range for the normal hierarchy case. For cases (i) and (ii) in the IH scenario, almost all the solutions lie outside the $3\sigma$ interval, while for case (iii) we now have the majority of the points within the same interval. Very few poits in the last case, though, do drop within the $1\sigma$ region, which show the robustness of the preference for NH of current neutrino oscillation data. The most noticeable effect of Majorana phases is the appearance of acceptable solutions in NH for the larger values of the neutrino scale.

\begin{figure}[h!]
  \centering
  \begin{subfigure}[b]{0.45\linewidth}
    \includegraphics[width=\linewidth]{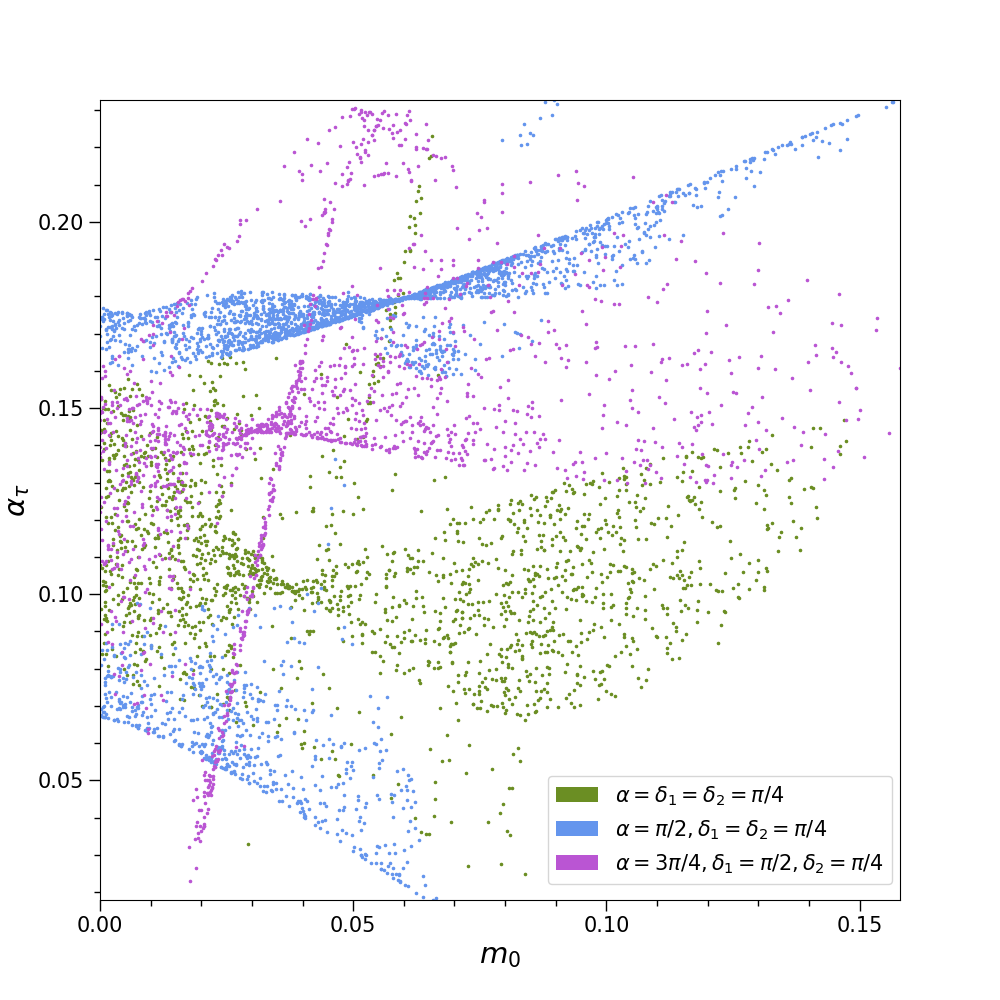}
    \caption{Normal hierarchy.}
  \end{subfigure}
  \begin{subfigure}[b]{0.45\linewidth}
    \includegraphics[width=\linewidth]{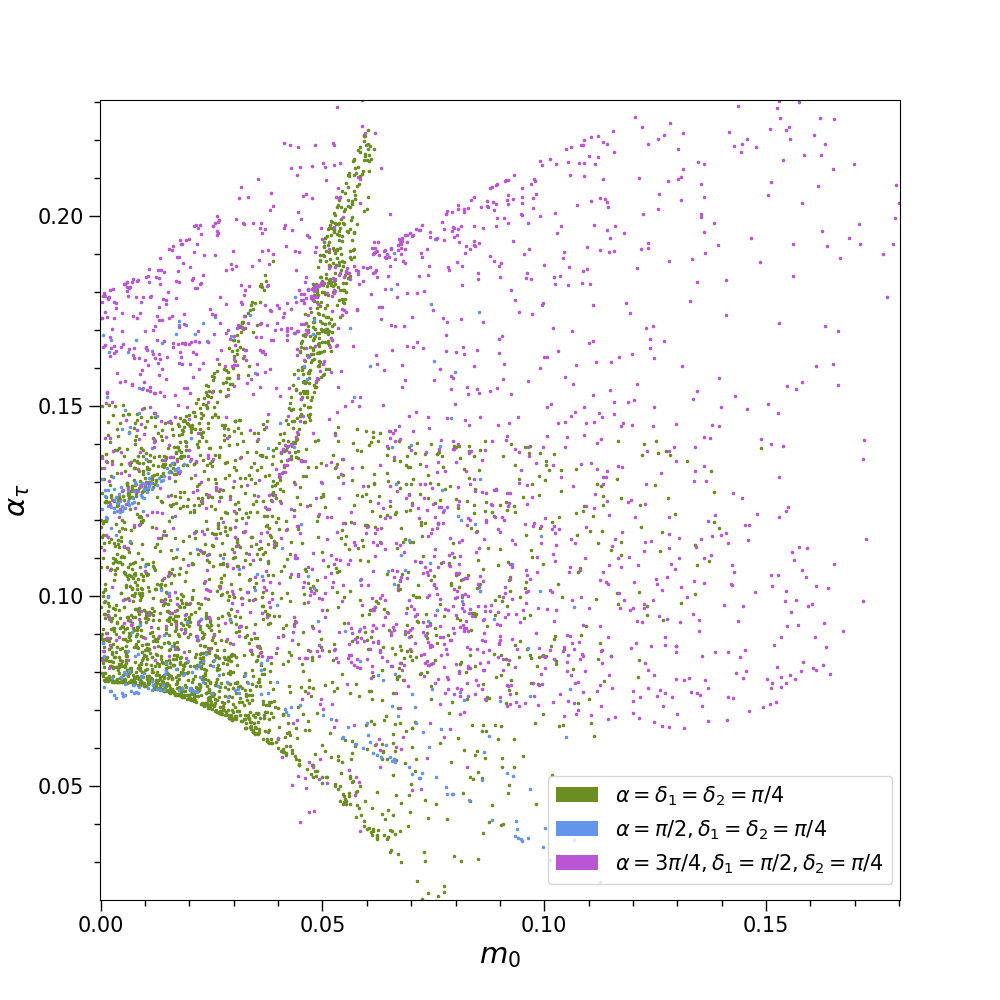}
    \caption{Inverted hierarchy.}
  \end{subfigure}
  \caption{Parameter space for $\alpha_\tau$, with non-zero Majorana phases, as a function of the lightest neutrino mass $m_0$.}
  \label{alpha_mp}
\end{figure}

In figure \ref{alpha_mp} we show the allowed values for $\alpha_\tau$ as a function of the mass $m_0$. For NH there is an approximately defined region for each of the cases (i-iii), while in the case of IH, the regions overlap between each other, nevertheless, this case would be of little phenomenological interest because the NH is preferred according to the present experimental data.

In the last two figures, the results for $\sin{I_2}$ and $\sin{I_3}$ are shown. Although these results seem quite arbitrary, they illustrate that the model has consistent solutions even when the Majorana phases are not null. 
Although these extra Dirac phases seem quite sensitive to the actual values of the Majorana phases, the presence of clear voids around zero values is noticeable. This indicates that for some ranges of the absolute neutrino scale, Dirac phases must definitively have non zero values.

\begin{figure}[h!]
  \centering
  \begin{subfigure}[b]{0.45\linewidth}
    \includegraphics[width=\linewidth]{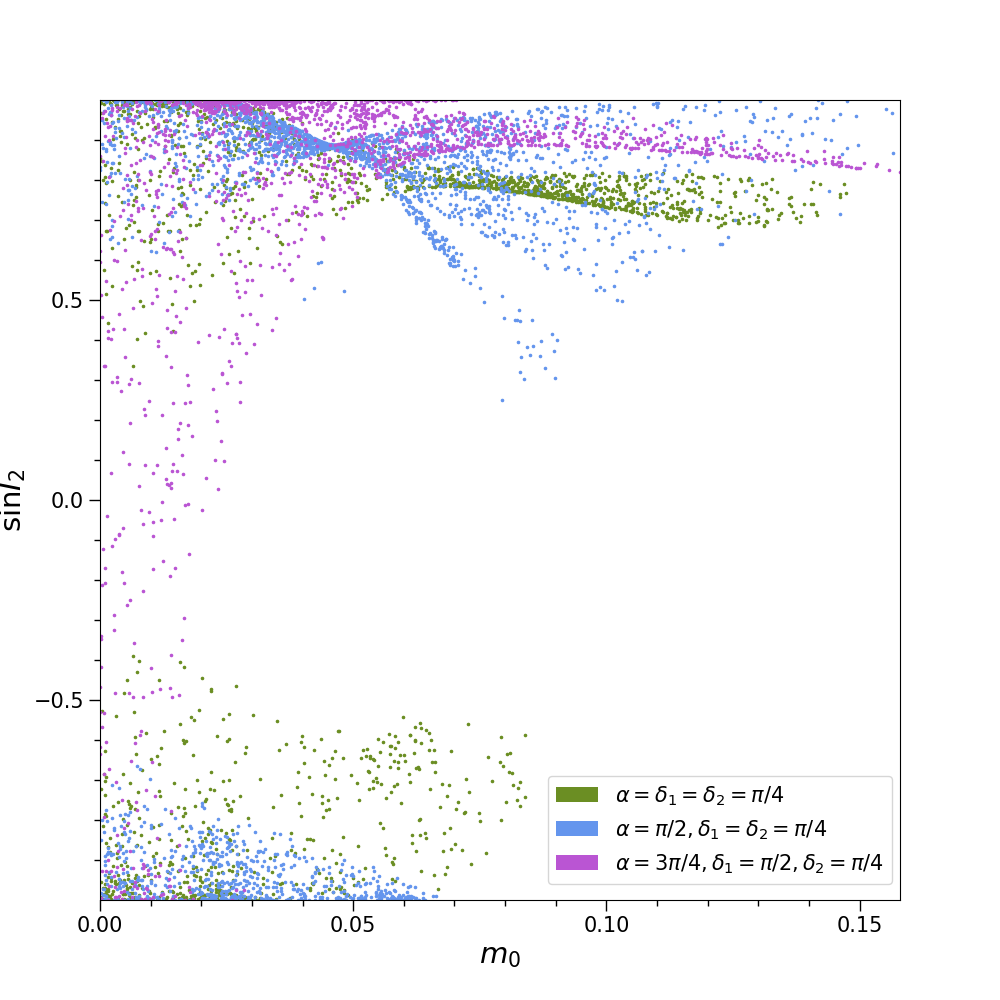}
    \caption{Normal hierarchy.}
  \end{subfigure}
  \begin{subfigure}[b]{0.45\linewidth}
    \includegraphics[width=\linewidth]{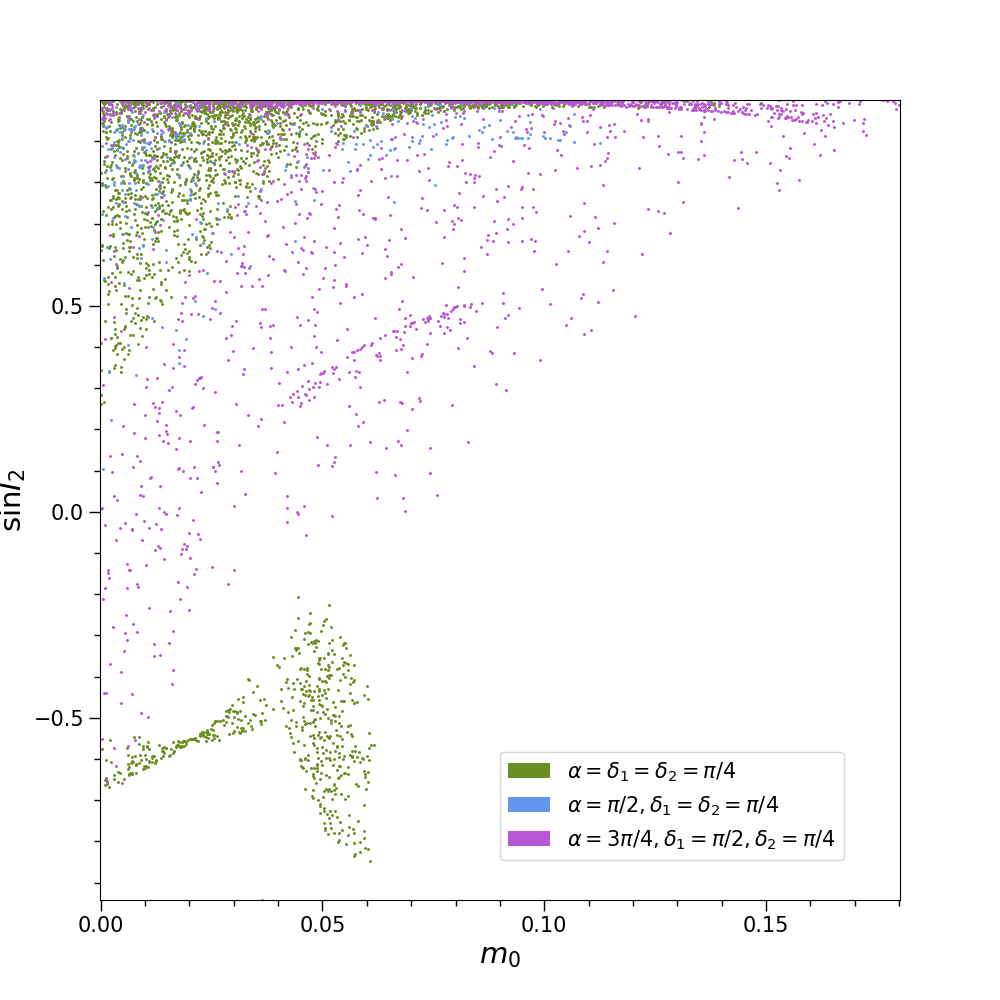}
    \caption{Inverted hierarchy.}
  \end{subfigure}
  \caption{$\sin{I_2}$ as a function of the lightest neutrino mass $m_0$ for non-zero Majorana phases.}
  \label{sinbeta_mp}
\end{figure}

\begin{figure}[h!]
  \centering
  \begin{subfigure}[b]{0.45\linewidth}
    \includegraphics[width=\linewidth]{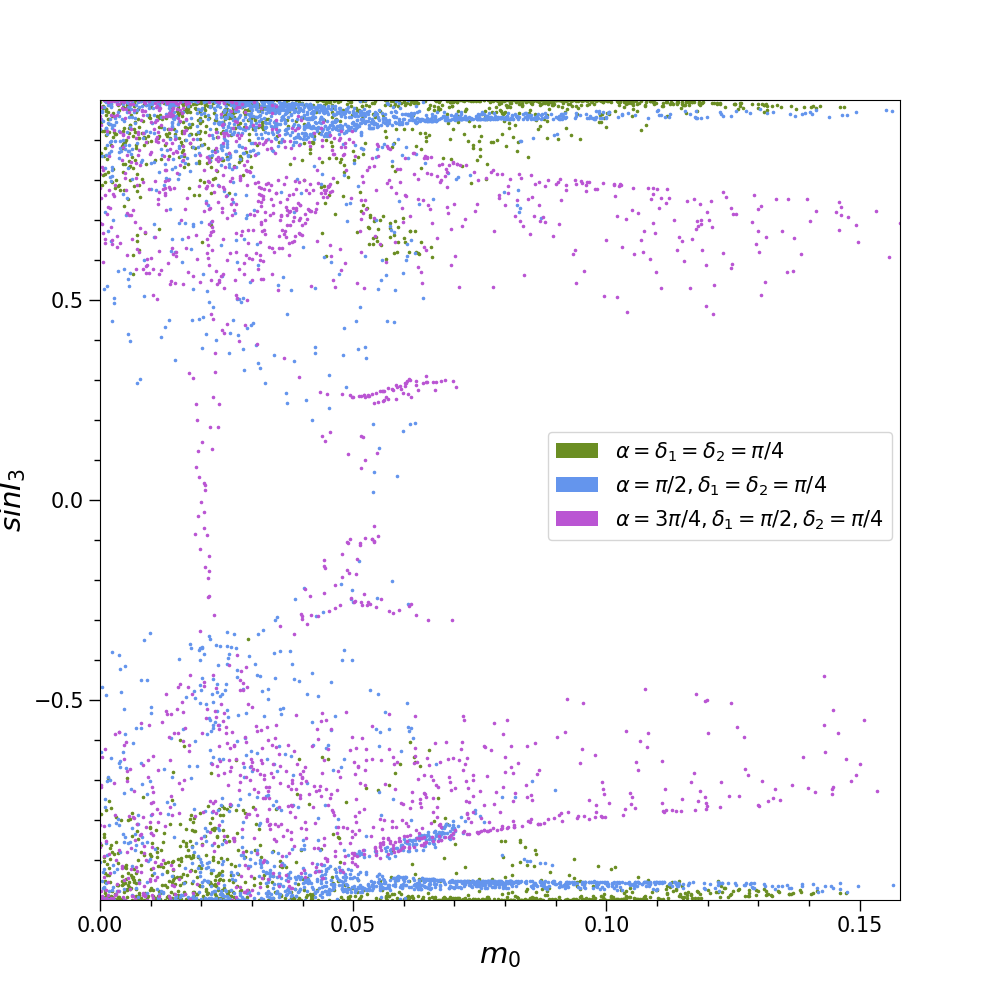}
    \caption{Normal hierarchy.}
  \end{subfigure}
  \begin{subfigure}[b]{0.45\linewidth}
    \includegraphics[width=\linewidth]{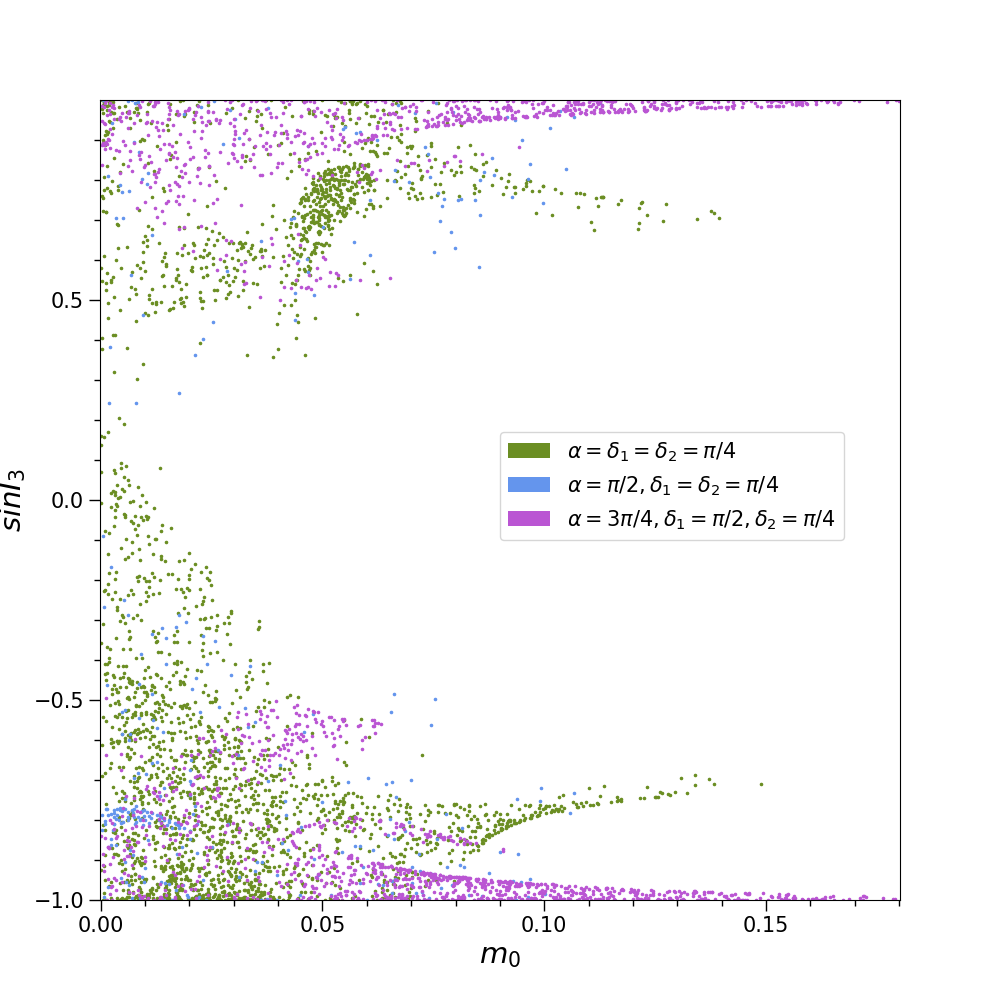}
    \caption{Inverted hierarchy.}
  \end{subfigure}
  \caption{$\sin{I_3}$ as a function of the lightest neutrino mass $m_0$ for non-zero Majorana phases.}
  \label{singamma_mp}
\end{figure}

\section{Concluding remarks}

To summarize, we have explored the parameter space of a $3+1$ scheme for neutrino oscillations, where exact $\mu-\tau$ symmetry is assumed to be valid in the active neutrino sector, but which is explicitly broken in the active-sterile neutrino mass matrix terms, searching for its predictions about  Dirac like phases.  By using the available neutrino data to constraint potential predictions, in the scenario where double beta decay amplitude is maximized, by assuming null Majorana like phases, we found that suitable solutions to the system exist that may accommodate a $\delta_{CP}$ (the Dirac phase appearing in the PMNS mixing matrix) well within current expected values within $1\sigma$ and $3\sigma$ deviations. Considering the decay of the sterile neutrino in the $3+1+decay$ scheme we found there is no a significant distinction for the parameters we explored as compared to the simpler $3+1$ model, hence, the results are robust enough to also fit the less constrained model in which the sterile neutrino is allowed to decay. It is noticeable that if actual $\delta_{CP}$ were found within the $1\sigma$ current interval, the normal hierarchy on active neutrino masses would be favored in the present model, since our explorations did not find any consistent region for such $\delta_{CP}$ for the inverted hierarchy. 
Moreover, our analysis also indicates that for such values of the active CP phase, a lightest neutrino mass  below $0.1~eV$, and thus a non-degenerate neutrino spectrum, is preferred.
It is also worth stressing the fact that the presented solutions on any hierarchy do predict that other Dirac like phases within the complete mixing matrix  to be non zero. 
Although we have centered our present discussion on the results obtained from  an specific set of values for $\alpha_e$ and $\alpha_\mu$ parameters, varying those does not actually add any new valuable information, as consistent solution are actually found for other value sets, in basically the same parameter regions. Switching on the Majorana phases shows the robustness of the NH scenario, incorporating consistent solutions that reproduce the experimental data even for the larger values of $m_0$, although the predicted parameter regions for the other Dirac phases may substantially change. Majorana phases may also bring back the IH scenario for the model at hand, provided future measurements of $\delta_{CP}$ keep values within the current $3\sigma$ region. 

Although the scenario at hand may still be seen as quite marginal, we believe it is still of interest as far it is not yet completely ruled out by data, as the addition of sterile neutrino decay has some potential of lessen the constraints on the sterile hypothesis. Given the potential to reproduce the observed neutrino oscillation parameters, it would be interesting to further explore complete flavor model constructions that may reproduce all these features. As already mentioned, earlier models were addressed in~\cite{Borah:2016fqj,Sarma:2018bgf,Das:2018qyt,Das:2019ntw} based on $A_4$ symmetry,  but  other discrete symmetries, as $S_3$ or $S_4$, could be also explored. A consistent introduction of heavy neutrino decay in the same flavor model basis is also a feature that remains to be fully explored.

\begin{acknowledgments}
This work has been partially supported by Conacyt, M\'exico, under FORDECYT-PRONACES Grant No. 490769. The work of E. Becerra-Garc\'ia was supported by a CONACyT graduate fellowship.
\end{acknowledgments}

\appendix
\section{Diagonalization of the neutrino mass matrix}

Here we outline the method for the diagonalization of the mass matrix, which is an extension of the one presented in \cite{King2002} for the four neutrino scenario. The diagonalization is achieved by the operation $U^T M_\nu U = M_D$ with $U$ given in (\ref{U_mix}) and considering the mixing angles $\theta_{i4}$ and $\theta_{13}$ to be sufficiently small for a perturbative analysis. In the following, $s_{ij}$ ($c_{ij}$) stands for $\sin{\theta_{ij}}$ ($\cos{\theta_{ij}}$).

\begin{enumerate}
    \item Perform the rotation $34$ as $\omega_{34}^T M_\nu \omega_{34}$, and set to zero the $34,43$ elements of the matrix, which allows us to obtain an expression for the mixing angle $\theta_{34}$, then we redefine the phases by demanding the angle to be real.
\begin{equation}
    \omega_{34}^T M_\nu \omega_{34}=
    \begin{pmatrix}
        \times & \times & \times & \times \hspace{1mm} \\
        \times & \times & \times & \times \hspace{1mm} \\
        \times & \times & \times & 0 \hspace{1mm} \\
        \times & \times & 0 & \times \hspace{1mm} \\
    \end{pmatrix},
\end{equation}
from the entry $34$ we obtain
\begin{equation}
    \theta_{34}=\frac{\alpha_\tau m_s e^{i\gamma}}{m_s e^{2i\gamma}-m_{\tau\tau}}
\end{equation}
    \item Now we apply the rotation $24$ to the resultant matrix of the previous step, and set to zero the entries $24,42$ after the operation, from which we obtain the mixing angle $\theta_{24}$, and again, in order for this angle to be real we need to redefine the phases. This second transformation produce a new quantity in the $34,43$ entries, but we neglect it because another rotation in the block $34$ to get rid of this term would produce contributions to second order in the small mixing angles and we are restricting the analysis to lower order. This argument also applies for the following transformation.
\begin{equation}
    \omega_{24}^T \omega_{34}^T M_\nu \omega_{34} \omega_{24}=
    \begin{pmatrix}
        \times & \times & \times & \times \hspace{1mm} \\
        \times & \times & \times & 0 \hspace{1mm} \\
        \times & \times & \times & 0 \hspace{1mm} \\
        \times & 0 & 0 & \times \hspace{1mm} \\
    \end{pmatrix},
\end{equation}
that set zeros in the $24$ and $42$ elements, so that we obtain the following expression, after proper phase redefinition,
\begin{equation}
    \theta_{24}=\frac{(m_{\mu\tau}\theta_{34} + \alpha_\mu m_s e^{i\gamma})e^{i(\beta-\gamma)}}{m_s e^{2i\beta}-m_{\mu\mu}}
\end{equation}
    \item Next, we make the third rotation with $14$, and set to zero the elements $14,41$ to obtain the angle $\theta_{14}$ after  redefining  the corresponding phases. 
\begin{equation}
    \omega_{14}^T \omega_{24}^T \omega_{34}^T M_\nu \omega_{34} \omega_{24} \omega_{14}=
    \begin{pmatrix}
        \times & \times & \times & 0 \hspace{1mm} \\
        \times & \times & \times & 0 \hspace{1mm} \\
        \times & \times & \times & 0 \hspace{1mm} \\
        0 & 0 & 0 & \times \hspace{1mm} \\
    \end{pmatrix}.
\end{equation}
The corresponding mixing angle is found to be
\begin{equation}
    \theta_{14}=\frac{(m_{e\mu}\theta_{24}e^{i(\alpha+\gamma)} + m_{e\tau}\theta_{34}e^{i(\alpha+\beta)}+ \alpha_e m_s e^{i(\alpha+\beta+\gamma)})e^{-i(\beta+\gamma)}}{m_s e^{2i\alpha}-m_{ee}}.
\end{equation}
    \item The resultant matrix of the last step has about null elements on the fourth column and fourth row, except of course on the entry $44$. Now, we should diagonalize the upper $3\times 3$ block corresponding to the active neutrino sector, starting with the $23$ rotation and proceeding as before with
\begin{equation}
    \omega_{23}^T\omega_{14}^T \omega_{24}^T \omega_{34}^T M_\nu \omega_{34} \omega_{24} \omega_{14} \omega_{23}=
    \begin{pmatrix}
        \times & \times & \times & 0 \hspace{1mm} \\
        \times & \times & 0 & 0 \hspace{1mm} \\
        \times & 0 & \times & 0 \hspace{1mm} \\
        0 & 0 & 0 & \times \hspace{1mm} \\
    \end{pmatrix}~.
\end{equation}
The equation for the $\theta_{23}$ angle then results into
\begin{equation}
    \tan{2\theta_{23}} = \frac{2(m_{\mu\tau} - \alpha_\mu m_s\theta_{34}e^{i\gamma})}{(m_{\tau\tau}-2 \alpha_\tau m_s\theta_{34}e^{i\gamma})e^{i\delta_3}-(m_{\mu\mu} - 2\alpha_\mu m_s\theta_{24}e^{i\beta})e^{-i\delta_3}}
\end{equation}
    \item The next rotation in the block $13$ sets to zero the $13,31$ entries to obtain the mixing angle $\theta_{13}$, after redefining the phases,
\begin{equation}
    \omega_{13}^T \omega_{23}^T \omega_{14}^T \omega_{24}^T \omega_{34}^T M_\nu \omega_{34} \omega_{24} \omega_{14} \omega_{23} \omega_{13}=
    \begin{pmatrix}
        \times & \times & 0 & 0 \hspace{1mm} \\
        \times & \times & 0 & 0 \hspace{1mm} \\
        0 & 0 & \times & 0 \hspace{1mm} \\
        0 & 0 & 0 & \times \hspace{1mm} \\
    \end{pmatrix}~.
\end{equation}
For the $\theta_{13}$ angle we obtain the following expression
\begin{equation}
    \theta_{13} = \frac{(m_{e\mu}s_{23}+m_{e\tau}c_{23}e^{i\delta_3}-\alpha_e m_s (\theta_{24}s_{23}e^{i\beta}+\theta_{34}c_{23}e^{i(\delta_3+\gamma)}))e^{i(\delta_2+\delta_3)}}{(m_{\mu\mu}s_{23}^2+m_{\tau\tau}c_{23}^2 e^{2i\delta_3}+m_{\mu\tau}s2_{23}e^{i\delta_3})e^{2i\delta_2}+(2\alpha_e m_s \theta_{14} e^{i\alpha}-m_{ee})e^{2i\delta_3}}
\end{equation}
    \item Finally, we perform the rotation $12$ and follow the same procedure as before. Thus, we get
\begin{equation}
    \tan{2\theta_{12}} = \frac{2(m_{12}-m_{14}\theta_{24}e^{i\beta})c_{23}-2(m_{13}-m_{14}\theta_{34}e^{i\gamma})e^{i\delta_3}s_{23}}{(k_1 c_{23}^2 + k_2e^{2i\delta_3}s_{23}^2 - k_3 e^{i\delta_3}s2_{23})e^{i\delta_1}-(m_{11}-2m_{14}\theta_{14}e^{i\alpha})e^{-i\delta_1}}
\end{equation}
where the quantities $k_1, k_2, k_3$ are given by
\begin{align}
    k_1 &= m_{22}-2m_{24}\theta_{24}e^{i\beta}, \\
    k_2 &= m_{33}-2m_{34}\theta_{34}e^{i\gamma}, \\
    k_3 &= m_{23}-m_{24}\theta_{34}e^{i\gamma}.
\end{align}
This completes the diagonalization process, with which we have obtained approximate expressions for all the mixing angles in terms of the elements of the mass matrix.

\end{enumerate}

\bibliography{DiracCPsneutrino-mutau.bib}

\end{document}